\documentclass[10pt,twocolumn,twoside]{IEEEtran} 

\usepackage{amsmath}
\usepackage{amsfonts}
\usepackage{amssymb}
\usepackage{mathrsfs}
\usepackage{dsfont}
\usepackage{bbm}
\usepackage{bm}
\usepackage{color}
\usepackage{hyperref}
\usepackage{booktabs}
\usepackage{graphicx}
\usepackage{subfigure}
\usepackage{caption}
\usepackage{epstopdf}
\usepackage{epsfig}
\usepackage{cite}
\usepackage{enumerate}
\usepackage{amsopn}
\usepackage{ntheorem}
\usepackage{algorithm,algpseudocode}

\newtheorem{theorem}{Theorem}
\newtheorem{remark}{Remark}
\newtheorem{lemma}{Lemma}

\newtheorem{corollary}{Corollary}

\newtheorem{problem}{Problem}

\DeclareMathOperator{\Tr}{Tr}
\DeclareMathOperator*{\argmin}{arg\,min}
\DeclareMathOperator*{\argmax}{arg\,max}

\makeatletter

\newcommand{\Rmnum}[1]{\expandafter\@slowromancap\romannumeral #1@}
\makeatother

\title{\LARGE \bf Learning Optimal Scheduling Policy for Remote State Estimation under Uncertain Channel Condition}

\author{Shuang Wu, Xiaoqiang Ren, Qing-Shan Jia, Karl Henrik Johansson, Ling Shi
\thanks{S. Wu and L. Shi are with the Department of Electronic and Computer Engineering, Hong Kong University of Science and Technology, Clear Water Bay, Kowloon, Hong Kong (e-mail: {swuak@ust.hk}, {eesling@ust.hk}).}
\thanks{X. Ren and K. H. Johansson are with EECS, KTH Royal Institute of Technology, Stockholm, Sweden (e-mail: {xiaoqren@kth.se}, {kallej@kth.se}).}
\thanks{Q.-S. Jia is with the Center for Intelligent and Networked Systems, Department of Automation, Beijing National Research Center for Information Science and Technology (BNRist), Tsinghua University, Beijing 100084, China (e-mail: {jiaqs@tsinghua.edu.cn})}
\thanks{The work of X. Ren and K. H. Johansson is supported in part by the Knut and Alice Wallenberg Foundation, the Swedish Strategic Research Foundation, and the Swedish Research Council.
}
\thanks{The work of Q.-S. Jia is supported in part by the National Key Research and Development Program of China (2016YFB0901900), the National Natural Science Foundation of China under grants (No. 61673229), and the 111 International Collaboration Project of China (No. B06002).}
}

\begin{document}

\maketitle

\begin{abstract}
We consider optimal sensor scheduling with unknown communication channel statistics. We formulate two types of scheduling problems with the communication rate being a soft or hard constraint, respectively. We first present some structural results on the optimal scheduling policy using dynamic programming and assuming that the channel statistics is known. We prove that the $Q$-factor is monotonic and submodular, which leads to threshold-like structures in both problems. Then we develop a stochastic approximation and parameter learning frameworks to deal with the two scheduling problems with unknown channel statistics. We utilize their structures to design specialized learning algorithms. We prove the convergence of these algorithms. Performance improvement compared with the standard $Q$-learning algorithm is shown through numerical examples, which also discuss an alternative method based on recursive estimation of the channel quality.
\end{abstract}


\section{Introduction}
The development of precision manufacturing enables massive production of small-sized wireless sensors. These sensors are deployed to collect data and transmit information for monitoring, feedback control and decision making~\cite{shuman2010measurement}. As the sensor nodes are often battery powered and the communication channel is shared by a large amount of devices, it is critical to optimize the transmission schedule of the sensors to systematically tradeoff the system performance with the sensor communication overhead~\cite{wang2006survey}.

In the last few decades, numerous studies have been dedicated to optimize the communication rate v.s. the estimation error of sensor nodes~\cite{imer2005optimal,lipsa2011remote,wu2013event,trimpe2014event,nourian2014optimal,chakravorty2017fundamental,ren2018infinite}. The general scheduling problems impose significant computation challenges due to its combinatorial nature. Sensor scheduling problems, however, usually possesses special structures and computation overhead can be reduced.  One common idea is that the sensor only transmits when the recently obtained information is important with respect to a certain criterion. For example, the work in~\cite{wu2013event,ren2018infinite} chose the criterion to be the certain norms of the innovation of a Kalman filter. The work in~\cite{trimpe2014event,nourian2014optimal} chose the criterion to be the variance of the estimation error.

The literatures in sensor scheduling can be categorized according to whether the underling communication channel is idealized~\cite{nourian2014optimal,chakravorty2017fundamental,ren2018infinite}, lossy~\cite{imer2005optimal,lipsa2011remote,wu2013event,trimpe2014event} or noisy~\cite{gao2018optimal}. The assumption of idealized channel condition ignores the underlying communication channel and simplifies the scheduling policy design. The design of the optimal transmission protocol for a non-ideal channel treats the channel as a part of the whole system and requires information of the communication channel conditions. The packet dropout process is often treated as Bernoulli process or a two-state Markov chain, while the channel noise is treated as an additive Gaussian white noise. Based on the channel model and its parameters, the optimal scheduling policy can be derived. However, acquiring information of the channel condition may be costly or even impossible~\cite{lau2006channel}. 

This paper considers optimal sensor scheduling over a packet-dropping channel with packet dropout rate unknown. We consider two scenarios. In the first scenario, the communication is costly. In the second scenario, there is an explicit communication rate constraint. We first prove monotonicity and submodularity of the $Q$-factor for these two types of problems, which leads to threshold-like structures in the optimal scheduling policy. We then design iterative algorithms to obtain the optimal solution without knowing the packet dropout rate. The major contribution of this work is as follows.
\begin{enumerate}
\item We show threshold-like structures (Theorems~\ref{theorem: threshold for costly comm} and~\ref{theorem: threshold for constrained comm}) of the optimal policy in the considered sensor scheduling problems. Specifically, the optimal policy for the costly communication problem (Problem 1) is a threshold policy and the optimal policy for the constrained communication problem (Problem 2) is a randomized threshold policy. These results are significant for scheduling problems as they leads to easy implementations and they have been reported in other papers under different setups (Discussions are in Section~\ref{sec: optiamal policy for known channel}. In this work, we further utilize these properties to improve the standard $Q$-learning algorithm.
\item We develop iterative algorithms based on stochastic approximation and parameter estimation, and compare them in the two different types of scheduling~problems. Based on the structure of the $Q$-factor, we devise structural learning methods, which impose the transient $Q$-factor to satisfy certain properties (Theorem~\ref{theorem: convergence of structured Q learning}). In addition, we develop a synchronous learning algorithm by utilizing the fact that the randomness of the state transition is independent of the particular state. By using that the optimal scheduling policy of the constrained communication problem can be written in a closed form, we show that an adaptive control method can be directly used to obtain the optimal scheduling policy (Theorem~\ref{theorem: convergence of the parameter learning}).
\end{enumerate}

In this work, we consider optimal scheduling with unknown channel conditions. We aim to adapt the scheduling policy to the real-time estimate of the channel condition. To yield an accurate estimate fo the channel condition, it is necessary to utilize the history of transmission successes and failures to determine the scheduling policy. An intuitive method is to compute the optimal scheduling policy based on the estimate of the channel condition, which is obtained by keeping track of some sufficient statistics of the channel state. By taking the scheduling decision as control actions and the remote state estimation error as system states, the optimal scheduling problem can be formulated as an optimal control problem. The computation of the optimal control law usually involves solving a Bellman optimality equation~\cite{puterman2005markov}, which is computationally intense. In this work, we develop iterative algorithms which are relatively easy to implement and reduce significant computation overhead compared with the intuitive method.

There are two main streams of research in the area of optimal control of unknown dynamic systems. One stream, termed as reinforcement learning~\cite{watkins1989learning,bertsekas1995neuro,sutton1998reinforcement}, combines the stochastic approximation and dynamic programming to iteratively solve the Bellman optimality equation. The basic idea is to iteratively ``learn" the value of each control decision at each state and take control actions based on the ``learned" values. The major drawback is that every state-action pairs are required to be visited comparably often so that the estimate of the value of the state-action pairs are accurate. The transient performance may not be desirable as suboptimal actions are taken to estimate the values.

The other stream uses an adaptive control approach which combines the parameter estimation and the optimal control. Under certain conditions, the ``certainty equivalence" holds, which implies a separation between parameter estimation and the optimal control. It is then optimal to take the parameter estimate as its actual value and take control actions based on the estimated parameters~\cite{kumar1986stochastic,hernandez1989adaptive,altman1991adaptive}. A major problem with the adaptive control is that computing the optimal control for a given parameter is computationally intense. We illustrate this with a numerical example in Section~\ref{sec: numerical example}. In this work, we utilize structures of the optimal policy to reduce the computation burden.

Both the reinforcement learning and the adaptive control frameworks guarantee that the iterative process converges to the optimal control policy under certain conditions. However, these works are quite generic. In specific problems, the special structure may be used to improve the transient performance. The sensor scheduling problem in this work possesses some structures in the optimal policy. We devise a learning scheme which takes advantage of these structures to improve transient performance and reduce computation overhead.

The remainder of this paper is organized as follows. In section~\Rmnum{2}, we provide the mathematical model of the sensor scheduling problem and two related optimization problems. In section~\Rmnum{3}, we use a dynamic programming approach to show structural results. In section~\Rmnum{4}, we present two learning frameworks to solve the optimal scheduling policy when the channel condition is unknown. We summarize the paper in section \Rmnum{5}. Proofs are given in the appendix.

\emph{Notations}: The bold symbol letter stands for a vector which aggregates all its components, e.g., $\boldsymbol{x}=[x_1,\dots,x_n]^{\top}$. For a matrix $X$, $\rho(X)$, $X^{\top}$ and $\Tr(X)$ stands for the spectral radius of the matrix, the matrix transpose and the trace of the matrix. The operation $[\boldsymbol{x}]_{\mathcal{X}}$ denotes the projection of vector $\boldsymbol{x}$ into the constrained set $\mathcal{X}$. The probability and the  conditional probability are denoted by $\mathrm{Pr}(\cdot)$ and $\mathrm{Pr}(\cdot | \cdot)$, respectively. The expectation of a random variable is $\mathbb{E}[\cdot]$. The set of nonnegative integers are represented by $\mathbb{N}$.

\section{Problem Setup}

\subsection{System Model}
\begin{figure}[t]
	\centering
	\includegraphics[width=0.45\textwidth]{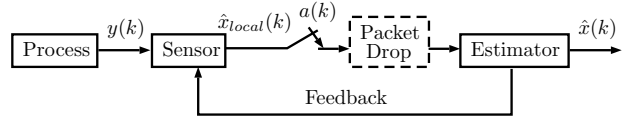}
	\caption{System architecture.}
	\label{fig:learning diagram}
\end{figure}

The architecture of the system is depicted in Fig.~\ref{fig:learning diagram}. We consider the following LTI process.
\begin{align*}
x(k+1) = Ax(k) + w(k),\\
y(k) = C{x}(k) + {v}(k),
\end{align*}
where ${x}(k)\in\mathbb{R}^{n}$ is the state of the process at time $k$ and ${y}(k)\in\mathbb{R}^{m}$ is the noisy measurement taken by the sensor. We assume, at each time $k$, that the state disturbance noise ${w}(k)$, the measurement noise ${v}(k)$, and the initial state ${x}(0)$ are mutually independent random variables, which follow Gaussian distributions as ${w}(k)\sim\mathcal{N}({0},{\Sigma_w})$, ${v}(k)\sim\mathcal{N}({0},{\Sigma_v})$, and ${x}(0)\sim\mathcal{N}({0}, {\Pi})$. We assume that the covariance matrices ${\Sigma_w}$ and ${\Pi}$ are positive semidefinite, and ${\Sigma_v}$ is positive definite. We assume that the pair $({A},{C})$ is detectable and that $({A},\sqrt{{\Sigma_w}})$ is stabilizable.

The sensor measures the process states and computes its local state estimates $\hat{x}_{local}(k)$ using a Kalman filter. After that, the sensor decides whether it should or not transmit the estimate through the packet-dropping communication channel to a remote state estimator. We use $a(k)=1$ to denote transmitting local estimate $\hat{x}_{local}(k+1)$ at time $k+1$ and $a(k)=0$ to denote no transmission. Let $\eta(k)=1$ denote that the packet is successfully received by the remote estimator at time $k$ and $\eta(k)=0$ otherwise. The successful transmissions are assumed to be independent and identically distributed as
\begin{align*}
\mathrm{Pr}(\eta(k+1)|a(k)=1)=
\begin{cases}
r_s,&\text{if }\eta(k+1)=1,\\
1-r_s,&\text{if }\eta(k+1)=0,\\
0,&\text{otherwise.}
\end{cases}
\end{align*}
Meanwhile, it is straightforward that $\mathrm{Pr}(\eta(k+1)=0|a(k)=0)=1$. 

The remote state estimator will either synchronize the remote state estimate with the local state estimate if the updated data is received, or use process dynamics to predict the state if no data is received. We assume that the local state estimate of the Kalman filter is in steady state. Define the remote state estimate as
\begin{align*}
\hat{x}(k)=\mathbb{E}[x(k)|\eta(0),\eta(0)\hat{x}_{local}(0),\dots,\eta(k),\eta(k)\hat{x}_{local}(k)].
\end{align*}
The mean square estimation error covariance of the remote estimator at time $k$, which is defined as
\begin{align*}
P(k)=\mathbb{E}&[(x(k)-\hat{x}(k))(x(k)-\hat{x}(k))^\top|\\
&\eta(0),\eta(0)\hat{x}_{local}(0),\dots,\eta(k),\eta(k)\hat{x}_{local}(k)],
\end{align*}
can be computed as follows:
\begin{align*}
P(k)=
\begin{cases}
\overline{P}, &\text{if~}\eta(k)=1,\\
AP(k)A^{\top}+\Sigma_w, &\text{if}~\eta(k)=0,
\end{cases}
\end{align*}
where $\overline{P}$ is the steady state of the state estimation error covariance of the Kalman filter.

\begin{figure}[t]
	\centering
	\includegraphics[width=0.4\textwidth]{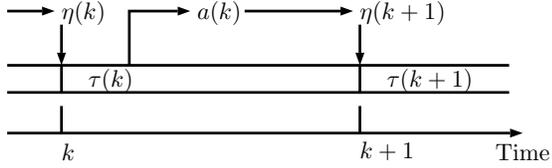}
	\caption{\noindent\textit{Relation among state $\tau(k)$, action $a(k)$ and transmission result $\eta(k)$.}}
	\label{fig: timestamp}
\end{figure}

The remote estimator will feed back a one-bit signal to the sensor to acknowledge its successful reception of the packet. The information of the remote state estimate available to the sensor for transmission decision is
\begin{align*}
\tau(k) = \min\{0\leq t \leq k: \eta(k-t)=1\},
\end{align*}
which is the time elapsed since the last successful transmission. The temporal relation among $a(k)$, $\eta(k)$ and $\tau(k)$ is illustrated in Fig.~\ref{fig: timestamp}.
Notice that $\tau(k)$ and $\eta(k)$ are equivalent in the sense that both of them can be used to compute the estimation error covariance at the remote estimator, which can be written as
\begin{align}\label{eq: error covariance and holding time}
&P(k)=\nonumber\\
&\begin{cases}
\overline{P},&\tau(k)=0,\\
A^{\tau(k)}\overline{P}(A^{\top})^{\tau(k)}+\sum_{t=0}^{\tau(k)-1}A^t\Sigma_w(A^{\top})^t,&\tau(k)\geq 1.
\end{cases}
\end{align}
An admissible scheduling policy $f=\{f_k\}_{k=0}^{\infty}$ is a sequence of mappings from $\tau_{0:k}$ and $a_{0:k-1}$ to the transmission decision, i.e.,
\begin{align*}
a(k)=f_k(\tau_{0:k},a_{0:k-1}),
\end{align*}
where $\tau_{0:k}$ and $a_{0:k-1}$ stand for $\tau(0),\dots,\tau(k)$ and $a(0),\dots,a(k-1)$, respectively. Denote $\mathbb{F}$ as the set of all admissible policies, i.e., policies that are measurable by $\tau_{0:k},a_{0:k-1}$.

\subsection{Performance Metrics and Problem Formulation}

Given a scheduling policy $f=\{f_k\}_{k=0}^{\infty}$, we define the expected average estimation error covariance of the remote estimator and the expected transmission rate. We use $\mathbb{E}^{f}$ to denote the expectation under the scheduling policy $f$. The expected average estimation error covariance is
\begin{align*}
J_e(f) = \limsup_{T\to\infty} \frac{1}{T} \mathbb{E}^{f} \Big[ \sum_{k=0}^{T-1} \Tr(P(k))) | P(0)=\overline{P}\Big],
\end{align*}
and the expected average transmission rate is
\begin{align*}
J_r(f) = \limsup_{T\to\infty} \frac{1}{T} \mathbb{E}^{f} \Big[ \sum_{k=0}^{T-1} a(k) | P(0)=\overline{P} \Big].
\end{align*}

We are interested in two optimization problems for these performance metrics.
\begin{problem}
	[Costly Communication] Given the communication cost for one transmission $\lambda$, solving the following minimization problem on the total cost:
	\begin{align*}
	\inf_{f\in\mathbb{F}} J_e(f)+\lambda J_r(f).
	\end{align*}
\end{problem}
\begin{problem}
	[Constrained Communication] Given a communication budget $b$, solving the following constrained minimization problem:
	\begin{align*}
	\inf_{f\in\mathbb{F}:J_r(f)\leq b} J_e(f).
	\end{align*}
\end{problem}

\begin{remark}\label{remark: relation between two problems}
The two problems are closely related. According to~\cite[Sec 11.4]{borkar2002convex}, if a policy $f^\star$ is a solution to Problem 2, then there exists a Lagrangian multiplier $\lambda^\star$ such that $f^\star$ minimizes $J_e(f)+\lambda^\star (J_r(f)-b)$, which means that $f^\star$ minimizes $J_e(f)+\lambda^\star J_r(f)$, i.e., $f^\star$ is a solution to Problem 1 with $\lambda=\lambda^\star$. However, even if $\lambda^\star$ is known beforehand, an optimal policy of Problem 1 may not be an optimal policy for the corresponding Problem 2. As it will later be shown, optimal policy of Problem 1 can be found in the set of deterministic policies while optimal policies of Problem 2 are randomized in general.
\end{remark}

We assume for the main results of this paper that the channel condition $r_s$ is unknown. When $r_s$ is known, Problems 1 and 2 can be solved via dynamic programming~\cite{chakravorty2017fundamental,ren2018infinite} or linear programming. Here, we cannot directly use the classical methods. We instead use a learning-based method. Dynamic programming approach is used to find some structures in the optimal scheduling policies. By utilizing the structures, we can accelerate the learning process.

A naive method to solve the problems is to iterate between estimating $r_s$ and solving the corresponding mathematical programming. However, the optimization problem then needs to be solved at each time step, which is computationally intense. In this work instead, we find a simple iterative method, which does not incur much computation overhead compared to the naive method.

\section{Optimal Scheduling Policy with Known Channel Condition}\label{sec: optiamal policy for known channel}
Before proceeding to the learning approach, we establish some structural results for the Problems 1 and 2 when assuming $r_s$ is known. We reformulate the original two problems using Markov decision process (MDP). The costly communication problem can be directly formulated as an MDP, while the constrained communication problem is a constrained MDP (coMDP). We will show the connection between these models.

Some of the results (e.g., Theorems~\ref{theorem: threshold for costly comm} and~\ref{theorem: threshold for constrained comm}) are similar to those in the literatures. The setup in~\cite{chakravorty2017fundamental,ren2018infinite,krishnamurthy2007structured} is different from ours. Leong et al.~\cite{leong2015optimality} showed the optimality of a threshold policy for the costly communication problem (Problem 1), but no results were developed for the constrained communication problem (Problem 2). In addition, they showed threshold property by studying the relative value function instead of the $Q$-factor as we do in this work.  To enable the structural learning procedure developed in the next section, we need to establish the monotonicity and submodularity of the $Q$-factor.

\subsection{Costly Communication}
An MDP $(\mathbb{S},\mathbb{A},\mathcal{P},c)$ consists of the state space $\mathbb{S}$, the action space $\mathbb{A}$, the state transition probability $\mathcal{P}$, and one stage cost $c$. In our formulation, the state space consists of all the possible $\tau(k)=\tau\in\mathbb{N}$. The action space consists of the transmission decision $a=a(k)\in\{0,1\}$. If action $a$ is taken when the current state is $\tau$, the state in the next time step will transit to $\tau_+$ according to the state transition probability
\begin{multline*}
\mathrm{Pr}(\tau_+|\tau,a)=
\begin{cases}
r_s, &\text{if }\tau_+=0 \text{ and } a=1,\\
1-r_s, &\text{if }\tau_+=\tau+1 \text{ and } a=1,\\
1, &\text{if }\tau_+=\tau+1 \text{ and } a=0,\\
0, &\text{otherwise.}
\end{cases}
\end{multline*}
The one stage cost is
\begin{align*}
c(\tau,a)=\Tr(P(\tau))+\lambda a,
\end{align*}
where we use $P(\tau)$ to emphasize that the estimation error can be determined by $\tau$ from~\eqref{eq: error covariance and holding time}. A policy corresponds to the scheduling policy $f:=\{f_k\}_{k=0}^\infty$, which maps the history $\tau_{0:k},a_{0:k-1}$ to the action space, i.e., $f_k(\tau_{0:k},a_{0:k-1})=a(k)$. By the Markovian property of the state transitions, it suffices to consider Markovian policies, the decision of which only depends on the current state. Therefore, we only need to consider the policies of the form of $a(k)=f_k(\tau(k))$.

The costly communication problem is compatible with the MDP model described above and its solution can be obtained by solving the following problem
\begin{align}\label{eq:solution to MDP}
\inf_{f\in \mathbb{F}^M} \limsup_{T\to\infty} \frac{1}{T+1} \mathbb{E} \Big[\sum_{k=0}^{T} c(\tau(k),a(k)) | \tau(0)=0 \Big],
\end{align}
where $\mathbb{F}^M$ is the set of all Markovian policies. Moreover, the optimal policy can be found in the set of all stationary policies $\mathbb{F}^S$, i.e., $\mathbb{F}^S=\{f:f_k=f_{k+1},\forall k\geq0\}$, if a stability condition holds.

\begin{lemma}\label{lemma: sen condition for acoe}
	If $\rho^2(A)(1-r_s)<1$, there exists a stationary policy $f^\star\in\mathbb{F}^S$ such that $a=f^\star(\tau)$ solves the Bellman optimality equation:
	\begin{multline}\label{eqe:bellman equation wrt value function}
	V(\tau) =\min_{a\in\mathbb{A}} \Big[ c(\tau,a) + \sum_{\tau_+}V(\tau_+)\mathrm{Pr}(\tau_+|\tau,a) - \mathcal{J}^\star \Big],
	\end{multline}
where $\mathcal{J}^\star$ is the optimal value of the trace of the average estimation error.
\end{lemma}

The stationary solution of the unconstrained MDP~\eqref{eq:solution to MDP} can be obtained by solving the Bellman optimality equation with respect to (w.r.t.) a constant $\mathcal{J}^\star$ and the relative value function $V(\tau)$. The optimal policy is to choose the action that minimizes the right hand side of~\eqref{eqe:bellman equation wrt value function}:
\begin{align*}
f(\tau)=\argmin_{a\in\mathbb{A}} \Big[ c(\tau,a) + \sum_{\tau_+}V(\tau_+)\mathrm{Pr}(\tau_+|\tau,a) - \mathcal{J}^\star \Big].
\end{align*}
We denote the value function of a state-action pair as 
\begin{align*}
Q(\tau,a)=c(\tau,a) + \sum_{\tau'}V(\tau')\mathrm{Pr}(\tau'|\tau,a)-\mathcal{J}^\star.
\end{align*}
Note that $V(\tau')=\min_{a\in\mathbb{A}} Q(\tau',a)$. We rewrite~\eqref{eqe:bellman equation wrt value function} as
\begin{align}\label{eq:bellman equation wrt q}
Q(\tau,a)=c(\tau,a) + \sum_{\tau'}\min_{u\in\mathbb{A}} Q(\tau',u)\mathrm{Pr}(\tau'|\tau,a)-\mathcal{J}^\star.
\end{align}

We can develop the following structural results for the $V$-function and the $Q$-factor.

\begin{lemma}[Monotonicity of $V(\cdot)$]\label{lemma: monotonicity of V}
	\begin{align*}
	V(\tau)\geq V(\tau'),~\forall \tau\geq \tau'.
	\end{align*}
\end{lemma}
\begin{lemma}[Monotonicity of $Q(\cdot,a)$]\label{lemma: monotonicity of Q}
	\begin{align*}
	Q(\tau,a)\geq Q(\tau',a), ~\forall \tau\geq \tau'.
	\end{align*}
\end{lemma}
\begin{lemma}[Submodularity of $Q(\cdot,\cdot)$]\label{lemma: submodularity of Q}
	\begin{align*}
	Q(\tau,a)-Q(\tau,a') \leq Q(\tau',a)-Q(\tau',a'), ~\forall \tau\geq \tau',~a\geq a'.
	\end{align*}
\end{lemma}

Thanks to monotonicity and submodularity, we have the threshold structure on the optimal policy for Problem 1\footnote{A similar result was also reported in~\cite{leong2015optimality}. We present it here for completeness and facilitate presentation of the structural learning as we utilized the monotonicity and submodularity of the $Q$-factor.}.

\begin{theorem}[Costly communication]\label{theorem: threshold for costly comm}
	The optimal policy $f^\star$ for Problem 1 with known channel condition $r_s$ is of threshold type, i.e., there exists a constant $\theta^\star\in\mathbb{S}$ such that
	\begin{align*}
	f^\star(\tau) =
	\begin{cases}
	0,&\text{if }\tau<\theta^\star,\\
	1,&\text{if }\tau\geq\theta^\star.
	\end{cases}
	\end{align*}
\end{theorem}

Since the optimal policy $f^\star$ is of threshold type, we use the threshold $\theta$ to represent a policy when there is no ambiguity. 

\begin{remark}
 Although similar results are available in literatures, either the setup is different~\cite{chakravorty2017fundamental,ren2018infinite,krishnamurthy2007structured}, or the results are obtained by imposing additional assumptions~\cite{ren2014dynamic}. Moreover, to our best knowledge, the structure of the $Q$-factor (monotonicity and submodularity) that is revealed in this work is the first of its kind in the field of sensor scheduling.
\end{remark}

\subsection{Constrained Communication}
The state space, action space and the transition probability of Problem 2 is the same as those of Problem 1. Nevertheless, two types of one stage cost are involved in the constrained communication problem
\begin{align*}
c_e(\tau,a)=\Tr(P(\tau))
\end{align*}
and
\begin{align*}
c_r(\tau,a)=a.
\end{align*}
Problem 2 can be formulated as a constrained MDP as
\begin{align*}
\inf_{f\in \mathbb{F}} \quad &\limsup_{T\to\infty} \frac{1}{T+1} \mathbb{E} \Big[\sum_{k=0}^{T} c_e(\tau(k),a(k)) | \tau(0)=0 \Big]\\
\text{s.t.} \quad &\limsup_{T\to\infty} \frac{1}{T+1} \mathbb{E} \Big[\sum_{k=0}^{T} c_r(\tau(k),a(k)) | \tau(0)=0 \Big] \leq b.
\end{align*}

We use the Lagrangian multiplier approach to convert the constrained problem to the following saddle point problem
\begin{align}
\inf_{f\in\mathbb{F}} \sup_{\lambda\geq 0} \limsup_{T\to\infty}\frac{1}{T+1}\mathbb{E}\Big[\sum_{k=0}^{T} c_e(\tau(k),a(k)) | \tau(0)=0 \Big] \nonumber\\
+ \lambda \Big( \limsup_{T\to\infty}\frac{1}{T+1}\mathbb{E}\Big[\sum_{k=0}^{T} c_r(\tau(k),a(k)) | \tau(0)=0 \Big] - b\Bigg).\label{eq:minmax constrained problem}
\end{align}

As the one-stage cost is bounded below and monotonically increasing, the above problem possesses a solution~\cite[Theorem 12.8]{altman1999constrained}. If we relax Problem 2 by fixing $\lambda$,~\eqref{eq:minmax constrained problem} reduces to Problem 1. Moreover, as the saddle point problem possesses a solution, there exists a $\lambda^\star$ such that the value of~\eqref{eq:minmax constrained problem} with $\lambda=\lambda^\star$ is the same as the value of the constrained problem (Remark~\ref{remark: relation between two problems}).

The following lemma constitutes a necessary condition for a policy to be optimal.
\begin{lemma}\label{lemma: necessary condition for optimal constrained policy}
If a scheduling policy $f\in\mathbb{F}$ solves Problem 2, it must satisfy
\begin{align*}
J_r(f)=b.
\end{align*}
\end{lemma}

From~\cite{altman1999constrained}, we know that as long as the constrained MDP is feasible, the optimal policy randomizes between at most $m+1$ deterministic policies, where $m$ is the number of constraints. Problem 1 has no constraints, the optimal policy is deterministic. Problem 2 has one constraint, so the optimal policy randomizes between at most two deterministic policies.

\begin{theorem}[Constrained communication]\label{theorem: threshold for constrained comm}
The optimal policy $f^\star$ for Problem 2 with known channel condition $r_s$ is of Bernoulli randomized threshold type, i.e., there exist two constants $\theta^\star\in\mathbb{S}$ and $0\leq r_{\theta^\star}\leq1$ such that
\begin{align*}
f^\star(\tau) =
\begin{cases}
0,&\text{ if }\tau<\theta^\star,\\
0,&with~probability~1-r_{\theta^\star}, \text{ if }\tau=\theta^\star,\\
1,&with~probability~r_{\theta^\star}, \text{ if }\tau=\theta^\star,\\
1,&\text{if }\tau>\theta^\star,
\end{cases}
\end{align*}
where $r_{\theta^\star}$ and $\theta^\star$ satisfy
\begin{align*}
\limsup_{T\to\infty} \frac{1}{T} \mathbb{E} \Big[   \sum_{k=0}^{T-1} f^\star(\tau(k)) \Big] = b.
\end{align*}
\end{theorem}

We see that the optimal policy for the Problem 2 only depends on the communication budget $b$ and the channel condition $r_s$. These relations are summarized in the following corollary.

\begin{corollary}\label{corollary: randomization policy}
The optimal threshold $\theta^\star$ for Problem 1 and randomization parameter $r_{\theta^\star}$ are given by
\begin{align*}
\theta^\star &= \lfloor \frac{1}{r_sb} -\frac{1}{r_s} \rfloor,\\
r_{\theta^\star} &=\theta^\star+1+\frac{b-1}{br_s},
\end{align*}
where $\lfloor\cdot\rfloor$ denotes the floor function.
\end{corollary}

\section{Optimal Scheduling Policy with Uncertain Channel Condition}
The optimal scheduling policy can be obtained if the $Q$-factor is solved by~\eqref{eq:bellman equation wrt q}. If the channel condition is not known beforehand, we cannot use classical solution techniques to solve the Bellman optimality equation. We propose two learning-based frameworks, stochastic approximation and parameter learning, to adaptively obtain the optimal policy without knowing the channel statistics a priori.

The stochastic approximation framework yields an iterative method to find a solution of the Bellman optimality equation. The optimal scheduling policy can be directly obtained from the $Q$-factor. The Bellman optimality equation~\eqref{eq:bellman equation wrt q} has a countable infinite state-space and cannot be solved directly. A finite-state approximation is needed. We restrict the largest state to be $M$, and any states larger than $M$ are treated as $M$. The optimal action on such states is to transmit the local estimate. As the optimal policy is a threshold-type, the optimal scheduling policy can be captured by solving a finite state approximation as long as $M$ is large enough. In other words, there exists an $M>0$ such that the optimal policy of any finite state approximation with $|\mathbb{S}|\geq M$ is the same as the optimal policy of the original model. In practice, we have to set a maximal interval between two transmissions for a sensor to avoid the sensor being always idle. The number $M$ can be set as the maximal interval. In the sequel, we denote $\mathbb{S}'$ as the truncated state space.

In parameter learning method, we continuously estimate the channel condition based on the scheduling results and compute the corresponding optimal scheduling policy by taking the estimated channel condition as the actual condition. As we have proven that the optimal policy for Problem 2 can be analytically computed, this method is more suitable for Problem 2.

In the following two sections, we discuss the stochastic approximation method for Problem 1 and 2. The parameter learning method is treated in a third section. Note that since the sensor is aware that whether a transmission succeeds through the feedback acknowledgment from the remote state estimator. The learning algorithm is thus done at the sensor.

\subsection{Problem 1 with Stochastic Approximation}
At each time step $k$, an action $a(k)$ is selected for state $\tau$ in an $\varepsilon$-greedy pattern as
\begin{align*}
a(k)=
\begin{cases}
\argmin_{u} Q_k(\tau,u),~&\text{with probaility}~1-\varepsilon,\\
\text{any action},~&\text{with probaility}~\varepsilon,
\end{cases}
\end{align*}
where $\varepsilon>0$ is a randomization parameter\footnote{The randomness is necessary because every state-action pairs should be visited with infinite number of times to guarantee convergence.} We then observe that the state transits to $\tau(k+1)=\tau'$. The iterative update of the $Q$-factor is
\begin{multline}
Q_{k+1}(\tau(k),a(k)) = Q_{k}(\tau(k),a(k))+\\ \alpha(\nu_k(\tau(k),a(k)))\Big[ c(\tau(k),a(k)) + \min_{u\in\mathbb{A}}Q_k(\tau(k+1),u) \\ - Q_k(\tau(k),a(k)) - Q_k(\tau_0,a_0) \Big]\label{eq: basic q learning_asynchronous},
\end{multline}
where $(\tau_0,a_0)$ is a fixed reference state-action pair, which can be arbitrarily chosen. The step size $\alpha(n)$ satisfies\footnote{Examples of such $\alpha(\cdot)>0$ include $1/n^p$ with $0.5<p\leq 1$, $\log(n)/n$ and $1/[n\log(n)]$.}
\begin{align*}
\sum_{n=0}^\infty \alpha(n)=\infty,\quad \sum_{n=0}^\infty [\alpha(n)]^2<\infty,
\end{align*}
and in~\eqref{eq: basic q learning_asynchronous} this step size depends on $\nu_k(\tau,a)=\sum_{n=0}^k\bm{1}[(\tau(n),a(n))=(\tau,a)]$, which is the number of times that the state-action pair $(\tau,a)$ has been visited.

The above scheme is proven to converge~\cite{abounadi2001learning}, but the convergence rate is slow in practice. One reason is that the scheme is asynchronous as only one state-action pair is updated at each time step. We propose two improvements for this scheme by updating as many state-action pairs as possible. We denote them structured learning and synchronous update.

\begin{remark}\label{remark: epsilon greedy}
The asynchronous algorithm does not converge to the actual $Q$-value under transition probability $\mathrm{Pr}(\tau_+|\tau,a)$ but a perturbed one as follows
\begin{align*}
\tilde{\mathrm{Pr}}(\tau_+|\tau,a) = (1-\varepsilon)\mathrm{Pr}(\tau_+|\tau,a) + \frac{\varepsilon}{|\mathbb{A}|}\sum_{u\in\mathbb{A}}\mathrm{Pr}(\tau_+|\tau,a).
\end{align*}
This scheme is suboptimal. A smaller $\varepsilon$ leads to a more accurate learning result but slows down the learning rate. The synchronous scheme, which will be introduced later, however, guarantees that the $Q$-value converges to its actual value as $\varepsilon$ can be set to zero.
\end{remark}

\begin{remark}
In addition to the randomization parameter $\varepsilon$, the truncation parameter $M$ and the stepsizes $\alpha$ also affects the learning process. A greater $M$ leads to a higher accuracy. As we mentioned in the beginning of this section, the communication rate of a sensor should  be above certain values. When $M$ is large enough so that the optimal threshold is below $M$, the size $M$ has very little effects on the accuracy. In light of transient behavior, big stepsizes lead to severe oscillation while small stepsizes lead to slow convergence rate. In practice, stepsizes of the form $\alpha(k) = \frac{c}{(1+k)^{a}}$, where $c$ is a constant and $0.5<a\leq1$, can be selected to tradeoff between fast convergence rate and small oscillations.
\end{remark}

\textbf{Structural learning}. The first improvement is based on the structural results proven in the previous section. We can infer the unvisited state-action pair by using the monotonicity and the submodularity structure on the $Q$-factor.  From this information, the $Q$-factor is closer to the solution of the Bellman optimality equation.

Submodularity of the $Q$-factor gives
\begin{align}\label{eq:neighor submodularity}
Q(\tau,1)-Q(\tau,0) -Q(\tau+1,1)+Q(\tau+1,0)\geq0, ~\tau\in\mathbb{S}'.
\end{align}
Stack the $Q$-factor for all state-action pair as a vector
\begin{align*}
\bm{Q} = \begin{bmatrix}
Q(0,0),&Q(0,1),&\dots,&Q(M,0),&Q(M,1)
\end{bmatrix}^\top
\end{align*}
We can then write~\eqref{eq:neighor submodularity} as
\begin{align*}
\bm{T}_s\bm{Q}\geq0,
\end{align*}
where 
\begin{align*}
\bm{T}_s = \begin{bmatrix}
&-1 &1 &1 &-1 &0 &0 &\dots \\
&0 &0 &-1 &1 &1 &-1 &\dots \\
& & & &\vdots \\
&\dots &0 &0 &-1 &1 &1 &-1 
\end{bmatrix}_{M \times 2(M+1)}
\end{align*}
and the inequality is performed element-wisely. Similarly, we can use the monotonicity constraint $Q(\tau+1,a)-Q(\tau,a)\geq0$ for all $\tau$ to write
\begin{align*}
\bm{T}_m\bm{Q}\geq0,
\end{align*}
where
\begin{align*}
\bm{T}_m = \begin{bmatrix}
&-1 &0 &1 &0 &0 &\dots \\
&0 &-1 &0 &1 &0 &\dots \\
& &  &\ddots & &\ddots & \\
& &\dots &0 &-1 &0 &1 
\end{bmatrix}_{2M \times 2(M+1)}
\end{align*}
The two constraints can be compactly written as $\bm{T}\bm{Q}\geq0$.

Suppose there is a function $g(\bm{Q})$ such that its gradient with respect to $\bm{Q}(\cdot,\cdot)$ fulfills
\begin{align}
\nabla_{\bm{Q}}g = &\Big[ c(\tau,a) + \sum \mathrm{Pr}(\tau'|\tau,a)\min_{u\in\mathbb{A}}\bm{Q}(\tau',u) \nonumber\\ &- \bm{Q}(\tau,a) - \bm{Q}(\tau_0,a_0) \Big]\label{eq: gradient of g}.
\end{align}
This iterative learning scheme is a gradient ascent algorithm for the maximization problem
\begin{align*}
\max_{\bm{Q}} \quad g(\bm{Q}).
\end{align*}

In the $Q$-learning algorithm, the expectation term in~\eqref{eq: gradient of g} is replaced with its noisy sample $\min_{u\in\mathbb{A}}Q_k(\tau(k+1),u)$.
We take the noisy sample of $(\tau(k),a(k))$ component of $\nabla_{\bm{Q}}g$, i.e., $\nabla_{\bm{Q}}g_{(\tau(k),a(k))}+N_k$, where
\begin{align*}
N(k)= &\Big[ c(\tau(k),a(k)) + \min_{u\in\mathbb{A}}\bm{Q}(\tau(k+1),u) \\
&- \bm{Q}(\tau(k),a(k)) - \bm{Q}(\tau_0,a_0) \Big]-\nabla_{\bm{Q}}g_{(\tau(k),a(k))}.
\end{align*}

Imposing the monotonicity and submodularity constraints on this optimization problem gives
\begin{align*}
\max_{\bm{Q}} \quad &g(\bm{Q})\\
\text{s.t.} \quad & \bm{T}\bm{Q}\geq0.
\end{align*}
For this problem, we consider the following primal-dual algorithm
\begin{align}
\bm{Q}_{k+1}&(\tau(k),a(k)) =  \bm{Q}_{k}(\tau(k),a(k))+\alpha(\nu(\tau(k),a(k))) \nonumber\\
&\times\Big[ \nabla_{\bm{Q}} g_{(\tau(k),a(k))} + N_k + [\bm{T}^{\top}\bm{\mu}_k]^{(\tau(k),a(k))} \Big],\label{eq:strutured learning primal}\\
\bm{\mu}_{k+1}& = \bm{\mu}_k- \alpha(k)\bm{T}\bm{Q}_k,\label{eq:strutured learning dual}
\end{align}
where $[\bm{T}^{\top}\bm{\mu}_k]^{(\tau,a)}$ corresponds to component $(\tau,a)$ of $\bm{T}^{\top}\bm{\mu}_k$. This algorithm converges to the solution of the Bellman optimality equation as stated in the following theorem.

\begin{theorem}\label{theorem: convergence of structured Q learning}
The structured $Q$-learning~\eqref{eq:strutured learning primal}-\eqref{eq:strutured learning dual} converges to a solution of~\eqref{eq:bellman equation wrt q} with probability $1$.
\end{theorem}

\begin{remark}
	Standard $Q$-learning uses the sample average to estimate the $Q$-factor. One sample is used to update one state-action pair. Our proposed method utilizes the monotonicity and submodularity of the $Q$-factor. This fully utilizes the samples, and potentially increases the convergence performance.
\end{remark}

\textbf{Synchronous update}. The second improvement is updating synchronously. In most cases, the synchronous update is not applicable for stochastic approximation-based real-time optimal control. In our problem, however, the randomness of the state transition is independent of the state. We can run a parallel virtual model with the actual model. The virtual model keeps track of the $Q$-factor and the actual model takes actions according to the $Q$-factor stored in the virtual model. Every time after the actual model transmits, we either observe successful transmission or failure. If the transmission is successful, the $Q$-factor is updated as
\begin{align}
Q_{k+1}(\tau,1) = &Q_{k}(\tau,1)+ \alpha\Bigg(\sum_{n=0}^k a(n)\Bigg)\Big[ c(\tau,1) + \min_{u\in\mathbb{A}}Q_k(0,u) \nonumber\\ &- Q_k(\tau,1) - Q_k(\tau_0,a_0) \Big],~\tau\in\mathbb{S}'.
\end{align}
If the transmission fails, the $Q$-factor is updated as
\begin{align}
&Q_{k+1}(\tau,1) =  Q_{k}(\tau,1)+ \alpha\Bigg(\sum_{n=0}^k a(n)\Bigg)\Big[ c(\tau,1) \nonumber\\ & + \min_{u\in\mathbb{A}}Q_k(\tau+1,u) - Q_k(\tau,1) - Q_k(\tau_0,a_0) \Big]~\tau\in\mathbb{S}'.
\end{align}
For $a=0$, the $Q$-factor is updated as
\begin{align}
&Q_{k+1}(\tau,0) =  Q_{k}(\tau,0)+ \alpha\Bigg(k - \sum_{n=0}^k a(n)\Bigg)\Big[ c(\tau,0) \nonumber\\ & + \min_{u\in\mathbb{A}}Q_k(\tau+1,u) - Q_k(\tau,0) - Q_k(\tau_0,a_0) \Big]~\tau\in\mathbb{S}'.
\end{align}
To summarize, the update of the $Q$-factor can be written as
\begin{align}
&Q_{k+1}(\tau,a) = Q_{k}(\tau,a)+ \alpha(i)\Big[ c(\tau,a) + \min_{u\in\mathbb{A}}Q_k(\tau',u) \nonumber\\ &- Q_k(\tau,a) - Q_k(\tau_0,a_0) \Big],~\tau\in\mathbb{S}',~a\in\mathbb{A},\label{eq:synchronous q learning}
\end{align}
where the next state $\tau'$ can be determined according to whether the transmission succeeds or not and the parameter $i$ in $\alpha(i)$ is
\begin{align*}
i = \begin{cases}
\sum_{n=0}^k a(n), ~\text{if}~a(k)=1,\\
k - \sum_{n=0}^k a(n), ~\text{if}~a(k)=0.
\end{cases}
\end{align*}
With this improvement, the randomness in the action selection is not necessary because every state-action pair is now updated simultaneously (Remark~\ref{remark: epsilon greedy}).

As the synchronous version is a standard $Q$-learning algorithm satisfying the assumptions made in~\cite{abounadi2001learning}, its convergence automatically holds.

\begin{remark}
The structural learning we introduce above can also be used for the synchronous version as the source of noise and the associated limiting ordinary differential equation (ODE) are the same.
\end{remark}

\begin{remark}
The randomization parameter $\varepsilon$ can be set to zero for the synchronous algorithm. Therefore, the $Q$-factor converges under the synchronous algorithm to the actual value of the model with original probability transition law $\mathrm{Pr}(\tau'|\tau,a)$. From the Bellman optimality equation, we can see that the average cost is a continuous function of the $Q$-factor. By the continuous mapping theorem~\cite[Theorem 3.2.4]{durrett2010probability}, the average cost also converges to the optimal one.
\end{remark}

\subsection{Problem 2 with Stochastic Approximation}
From the structural results for Problem 2, we know that, for each communication budget $b$, there exists a $\lambda^\star(b)$ such that the optimal total cost with communication cost being $\lambda^\star(b)$ for Problem 1 equals to the optimal average estimation error under communication budget $b$ plus $\lambda^\star(b)b$. We use a gradient-based update for the communication cost to obtain $\lambda^\star(b)$ as follows
\begin{align}
\lambda_{k+1}=&\lambda_k+\beta(k)\Big( a(n)-b\Big),\label{eq:price learning}
\end{align}
where $\beta(k)$ is the step size at time $k$. From previous analysis, we know that the optimal randomized policy for Problem 2 is also an optimal policy for Problem 1 with communication cost being $\lambda^\star(b)$.

Combing~\eqref{eq:synchronous q learning}-\eqref{eq:price learning}\footnote{Such combination is also applicable for the original asynchronous version and the structural learning. Convergence analysis of these are the same.}, the iterative learning algorithm for the Problem 2 is
\begin{align*}
Q_{k+1}(\tau,a) = &Q_{k}(\tau,a)+ \alpha[\sum_{n=0}^k a(n)]\Big[ c_{\lambda_k}(\tau,a) + \min_{u\in\mathbb{A}}Q_k(\tau',u) \\ &- Q_k(\tau,a) - Q_k(\tau_0,a_0) \Big],~\tau\in\mathbb{S}',~a\in\mathbb{A},\\
\lambda_{k+1}=&\lambda_k+\beta(k)\Big( a(k)-b\Big),
\end{align*}
where the subscript $\lambda_k$ in $c_{\lambda_k}(\cdot,\cdot)$ is used to emphasize the dependence of the one stage cost on the communication cost. The step sizes $\alpha(\cdot)$ and $\beta(\cdot)$ satisfy
\begin{align}
&\sum_{n} \alpha(n)=\sum_{n} \beta(n) = \infty,~\sum_n (\alpha(n))^2+(\beta(n))^2<\infty,\label{eq:step size constraint multitime}\\
&\text{and}~\lim_{n\to\infty} \frac{\beta(n)}{\alpha(n)} = 0.\label{eq:step size constraint multitime quasi-static}
\end{align}

The last requirement imposes that that the communication cost $\lambda$ is updated in a slower time scale. This is called a quasi-static condition because the updates of $\lambda$ seem ``static" when $\bm{Q}$ is updating. By using either the standard asynchronous $Q$-learning or its improved version discussed before, for every ``static" cost $\lambda$, the vector $\bm{Q}$ converges to the corresponding solution of the Bellman optimality equation~\eqref{eq:bellman equation wrt q}. Consequently, the scheduling policy will also converge to the optimal one. If the algorithm over the slower time scale also converges, the two-time scale algorithm converges. This result is stated in the following theorem.

\begin{theorem}\label{theorem: convergence of the two-time scale}
The two-time scale $Q$-learning~\eqref{eq:synchronous q learning}-\eqref{eq:price learning} converges with probability 1. The asymptotic communication cost $\lambda_{\infty}=\lambda^\star$ and the $Q$-factor are the solutions to the Bellman optimality equation
\begin{align*}
Q(\tau,a)=c(\tau,a) + \sum_{\tau_+}\min_{a\in\mathbb{A}} Q(\tau_+,a)\mathrm{Pr}(\tau_+|\tau,a)-\mathcal{J}^\star,
\end{align*}
with $c(\tau,a)=\Tr(P(\tau))+\lambda^\star a$. The optimal policy $f(\lambda^\star)$ satisfies $J_r(f(\lambda^\star))=b$.
\end{theorem}

\subsection{Problem 1 and Problem 2 with Parameter Learning}
The stochastic approximation method iteratively updates the $Q$-factor. In the sensor scheduling problem, only the transmission success probability is unknown. If we can sample the channel condition infinitely many times, the empirical success probability converges to the actual success probability almost surely by the strong law of large numbers. Based on this observation, we develop the direct learning schemes for Problems 1 and 2, respectively. Different from previous sections, we discuss Problem 2 first.

\subsubsection{Problem 2}
Thanks to Theorem~\ref{theorem: threshold for constrained comm}, the optimal policy in this case only depends on the channel condition $r_s$ and the communication budget $b$. Once we know the channel condition $r_s$, the optimal threshold $\theta^\star$ and switching probability $r_{\theta^\star}$ can be analytically computed as shown in Corollary~\ref{corollary: randomization policy}. We propose the following learning method. Let $N_s(k)$ and $N_f(k)$ denote the number of successful transmission and failed transmission at time $k$. The maximum likelihood estimate of $r_s$ is
\begin{align}
\hat{r}_s(k)=\frac{N_s(k)}{N_s(k)+N_f(k)}.\label{eq:rate estimate}
\end{align}
We use $\hat{r}_s$ instead of $r_s$ to determine the corresponding optimal scheduling policy as
\begin{align}
\theta^{\star}(\hat{r}_s(k)) &= \lfloor \frac{1}{\hat{r}_s(k)b} -\frac{1}{\hat{r}_s(k)} \rfloor,\label{eq:threshold estimate}\\
r_{\theta^{\star}}(\hat{r}_s(k)) &=\theta^{\star}(\hat{r}_s(k))+1+\frac{b-1}{b\hat{r}_s(k)}\label{eq:bernouli estimate}.
\end{align}
If $\hat{r}_s=0$, the corresponding threshold is defined to be infinity. This can be avoided through proper initialization. In the initialization phase, we keep transmitting until $N_s(k)=1$. After that, we use the randomized threshold policy $(\theta^{\star}(\hat{r}_s(k)),r_{\theta^{\star}}(\hat{r}_s(k)))$ to determine the scheduling policy while learning $r_s$.

This scheme separates the parameter estimation and the optimal control problem. Its convergence is immediate.
\begin{theorem}\label{theorem: convergence of the parameter learning}
The schedule policy, which uses~\eqref{eq:rate estimate}-\eqref{eq:bernouli estimate} converges almost surely to the optimal policy. Moreover, also the average estimation error converges to the optimal average estimation error almost surely.
\end{theorem}

\subsubsection{Problem 1}
In this case, the estimation of $r_s$ and its initialization remains the same as in Problem 2. As shown before, there is no analytic expression for the optimal policy in Problem 1. For every given channel condition estimate $\hat{r}_s(k)$, we need to solve the Bellman optimality equation. As the initialization guarantees that $\hat{r}_s(k)$ will not be zero, the corresponding policy is a finite-threshold policy, which ensures that the trial does not stop. However, the computation overhead is large as the Bellman optimality equation needs to be solved at each time step. We present a numerical example to illustrate the computational issue in this scenario.

\begin{remark}
To summarize this section, the parameter learning method is suitable for Problem 2. Meanwhile, the stochastic approximation causes less computation overhead than the parameter learning method for Problem 1.
\end{remark}

%
%
%
%
\begin{figure}[t]
	\subfigure[Asynchronous algorithm.]{
		\centering
		\includegraphics[width=0.22\textwidth]{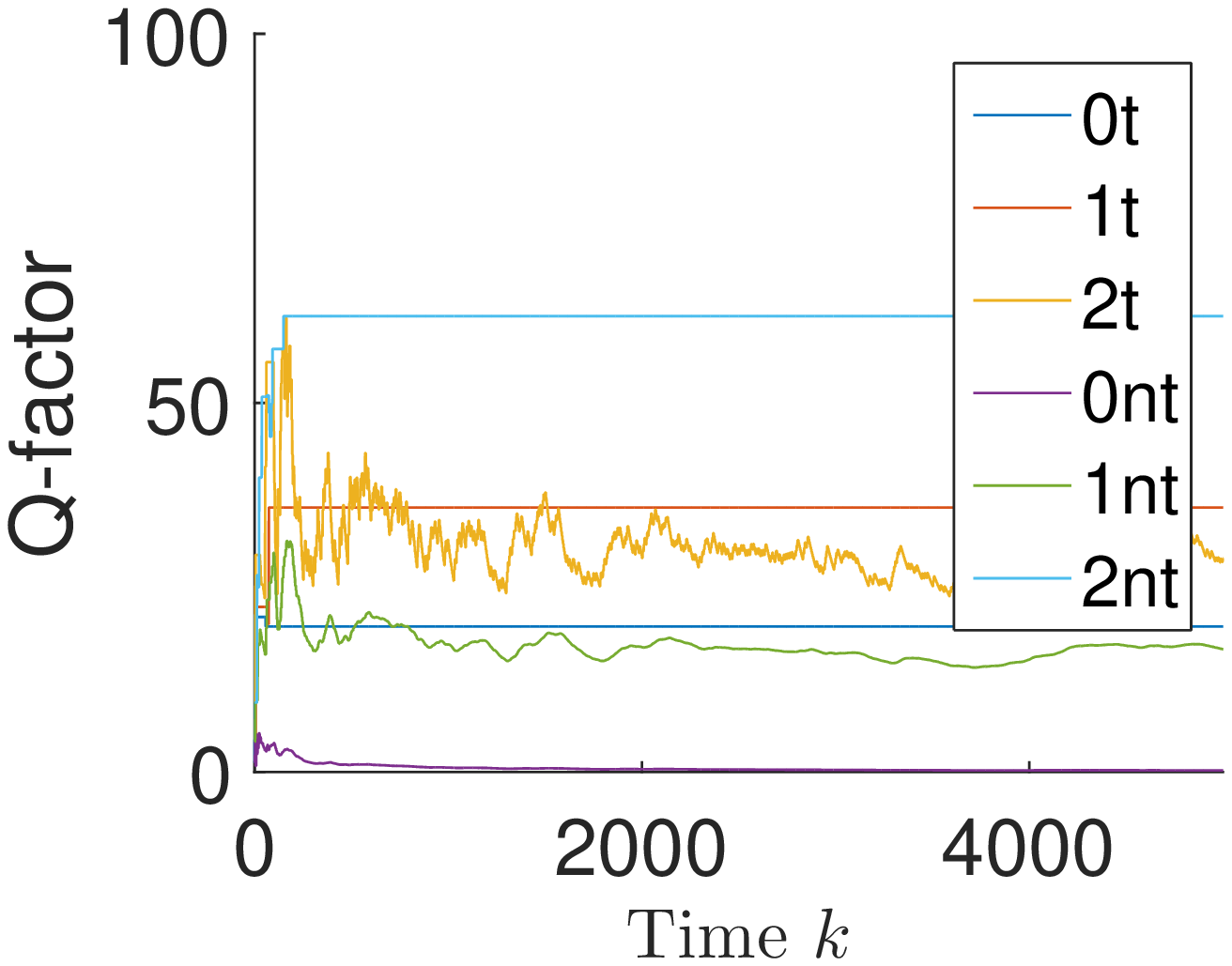}
	}
	\hspace{1em}
	\subfigure[Strctured asynchronous algorithm.]{
		\centering
		\includegraphics[width=0.22\textwidth]{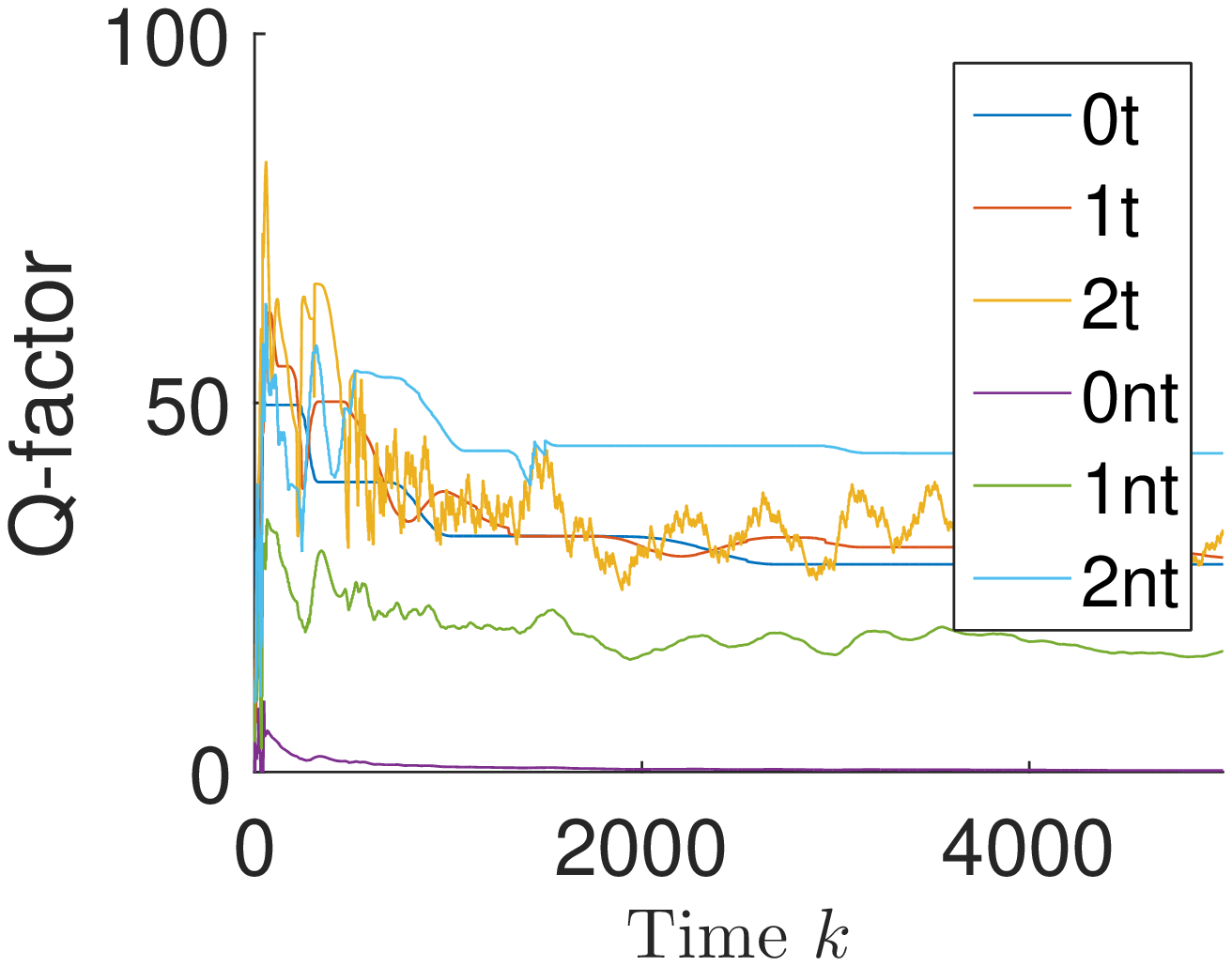}
	}
	\subfigure[Synchronous algorithm.]{
		\centering
		\includegraphics[width=0.22\textwidth]{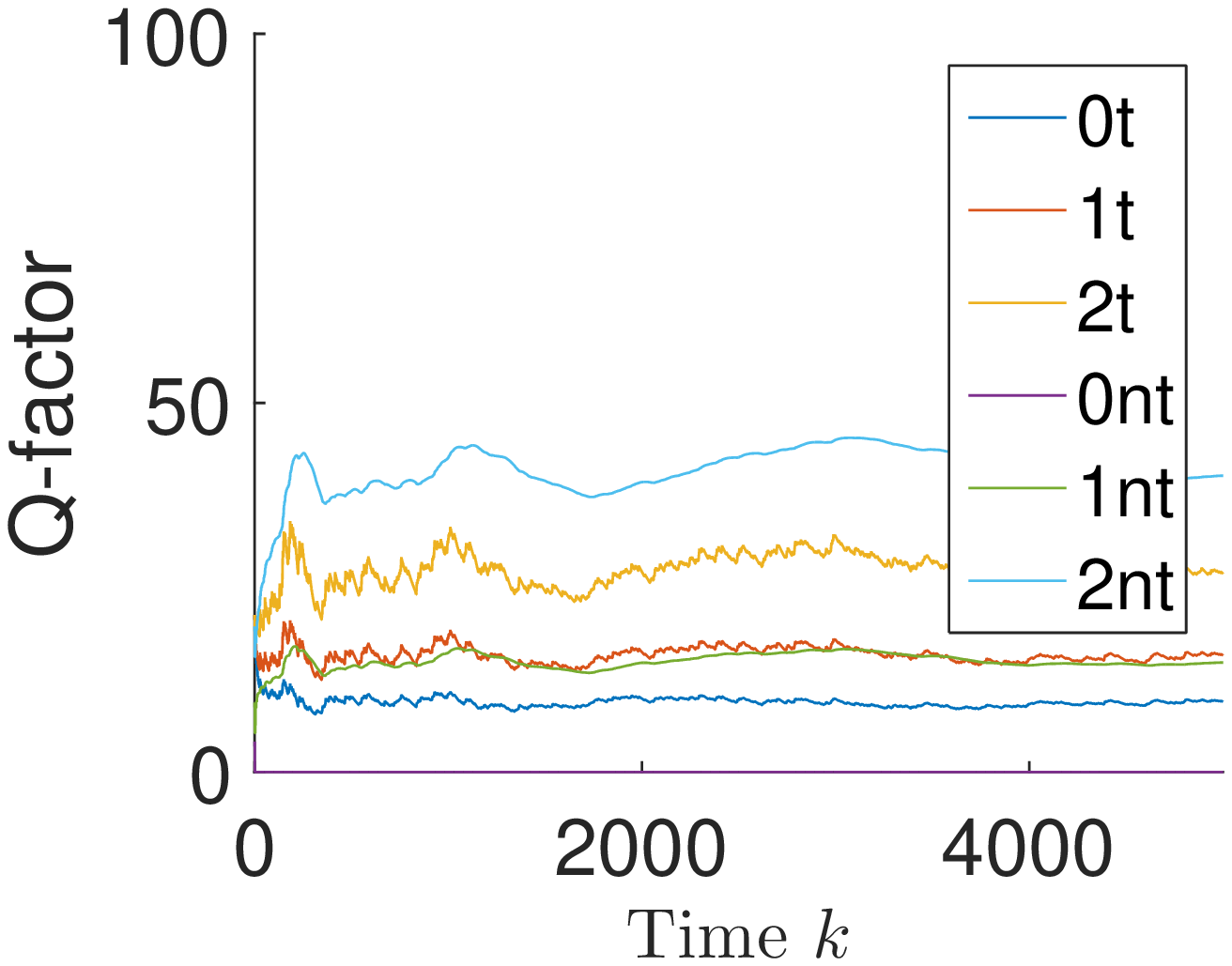}
	}
	\hspace{1em}
	\subfigure[Strucutred synchronous algorithm.]{
		\centering
		\includegraphics[width=0.22\textwidth]{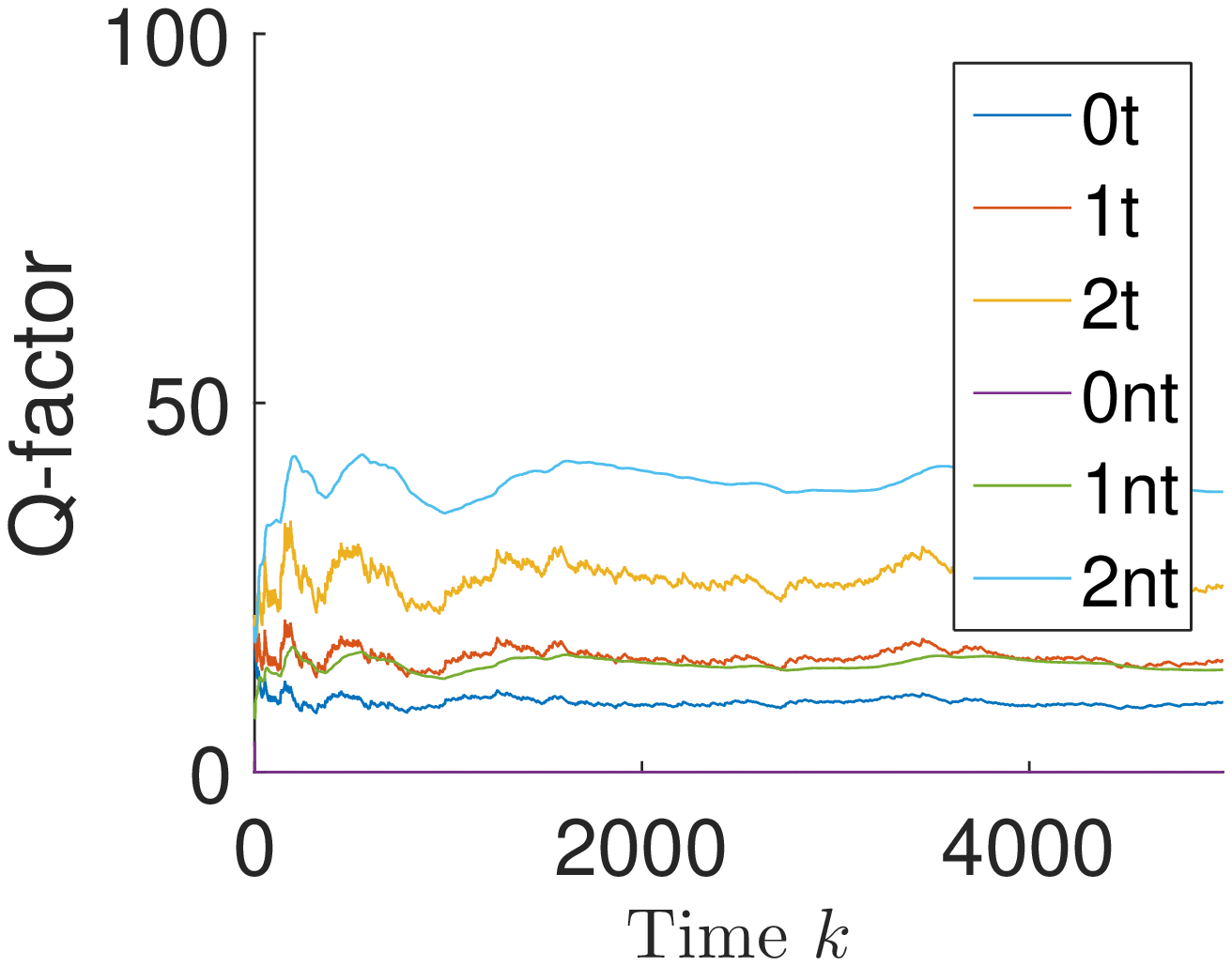}
	}
	\caption{$Q$-factor in the learning process for Problem 1.}
	\label{fig:Q}
\end{figure}
\section{Numerical Example}\label{sec: numerical example}
In this section, we illustrate the convergence of our algorithms with a specific example. We consider the following system:
\begin{align*}
x(k+1) &= \begin{bmatrix}1.2 &1 \\ 0 &0.8\end{bmatrix}x(k) + w(k),\\
y(k) &= {x}(k) + {v}(k),
\end{align*}
where
\begin{align*}
\mathbb{E}[w(k)w(k)^\top]=\begin{bmatrix}1 &0 \\ 0 &1\end{bmatrix},\mathbb{E}[v(k)v(k)^\top]=\begin{bmatrix}1 &0 \\ 0 &1\end{bmatrix}.
\end{align*}
The successful transmission rate is $r_s=0.7$.
\begin{figure}[t]
	\centering
	\includegraphics[width=0.3\textwidth]{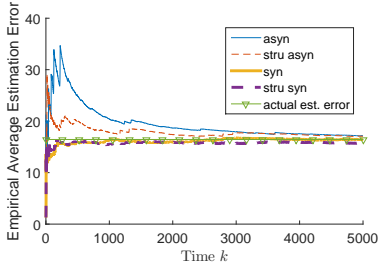}
	\caption{Average estimation error in Problem 1.}
	\label{fig:costly_aver_cost}
\end{figure}
\begin{figure}[t]
	\centering
	\includegraphics[width=0.3\textwidth]{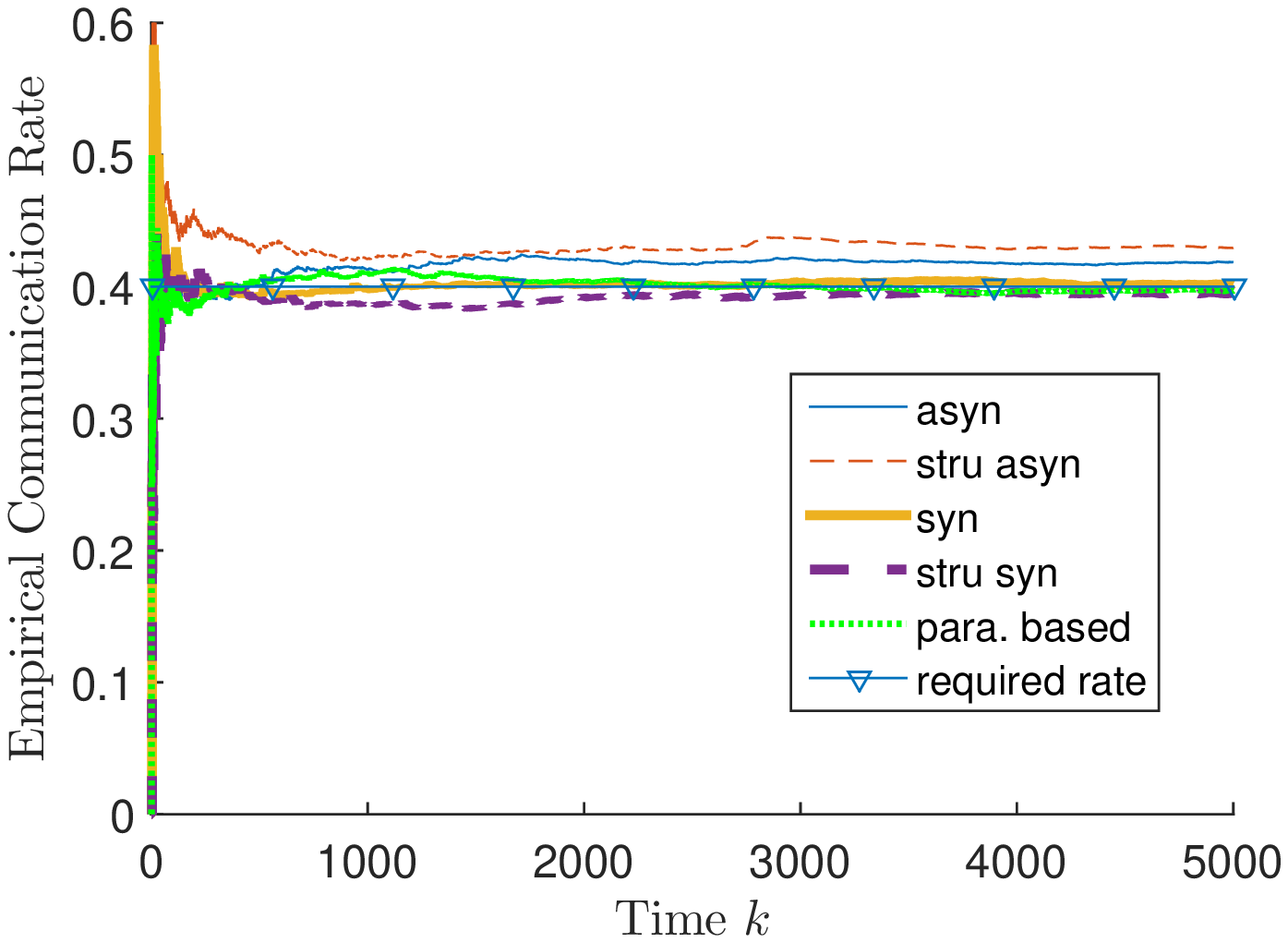}
	\caption{Communication rate in Problem 2.}
	\label{fig:constrained_comm_rate}
	\includegraphics[width=0.3\textwidth]{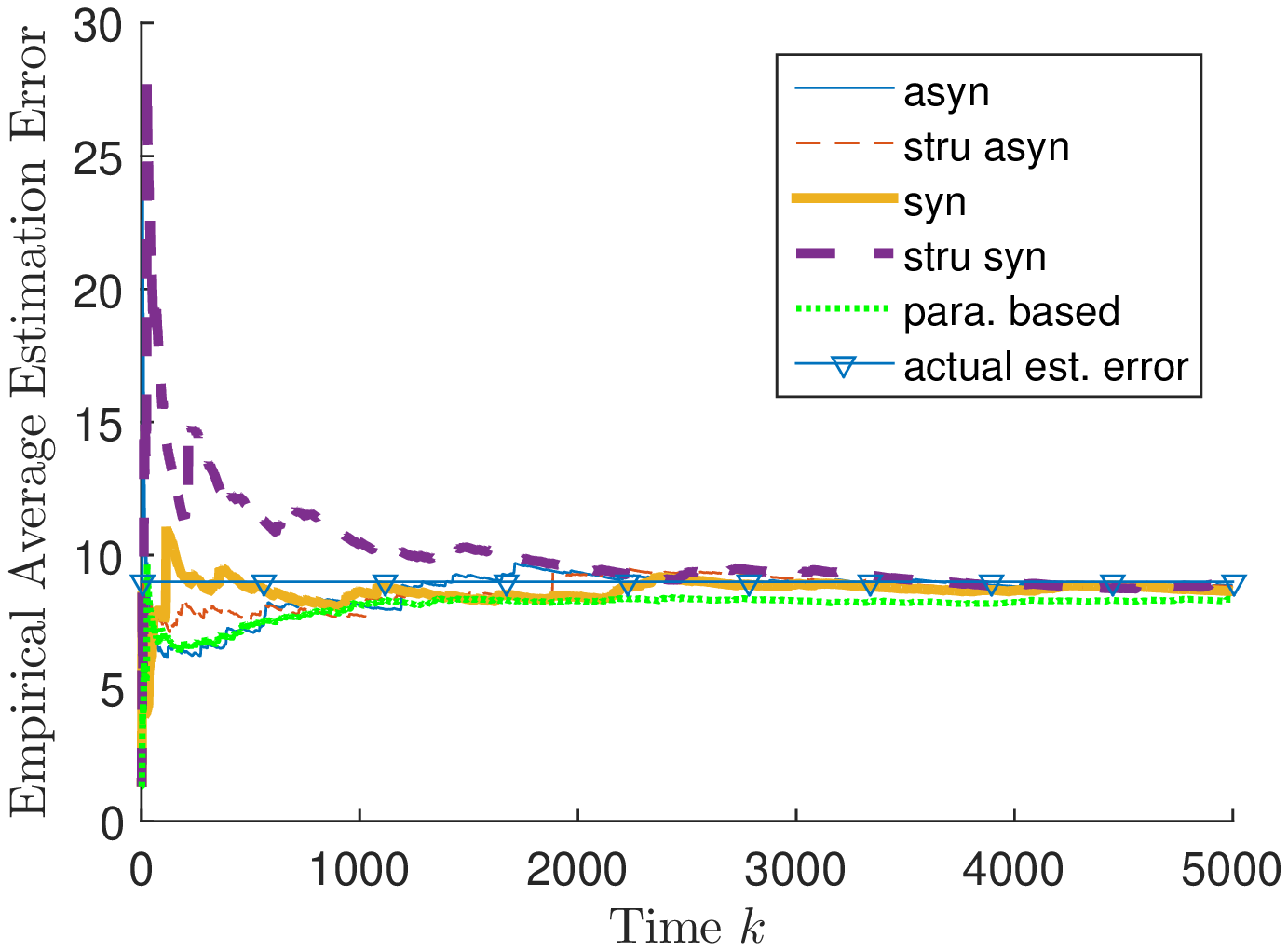}
	\caption{Average estimation error in Problem 2.}
	\label{fig:constrained_aver_cost}
\end{figure}

We first consider the Problem 1. We set the communication cost to be $\lambda = 20$ per transmission. We compare four algorithms: the original asynchronous algorithm~\eqref{eq: basic q learning_asynchronous}, the structure-based asynchronous algorithm in~\eqref{eq:strutured learning primal}-\eqref{eq:strutured learning dual}, the synchronous algorithm~\eqref{eq:synchronous q learning} and the structure-based synchronous algorithm (a combination of the structural learning and the synchronous algorithm). The learning processes of all algorithms converge as shown in Fig.~\ref{fig:Q}. The label $n$t ($n$nt) stands for transmit (not transmit) when $\tau=n$. We can see that the $Q$-factor in the original asynchronous algorithm does not satisfy the monotonicity condition. By comparing (a) and (b), we can see that the structure-based learning ensures monotonicity and submodularity of the $Q$-factor. The average cost, which is the empirical sum of the time average of the estimation error and the average communication cost, is shown in Fig.~\ref{fig:costly_aver_cost}. Specifically, the empirical estimation error at time $k$ is computed as
\begin{align*}
\tilde{J}_e(k) =\frac{1}{k+1} \sum_{t=0}^{k} c_e(\tau(t),a(t))
\end{align*}
and the empirical communication rate as
\begin{align*}
\tilde{J}_r(k) =\frac{1}{k+1} \sum_{t=0}^{k} a(t).
\end{align*}
For comparison, we also provide the true value of cost. We can see that all four algorithms converge to the true value. As the structure-based learning imposes the monotonicity and the submodularity of the $Q$-factor, the average estimation error in the structure-based asynchronous version converges to the true value faster than the basic asynchronous version. Moreover, the synchronous algorithms have much faster convergence rate than the asynchronous ones as expected.

We then consider the Problem 2. We set the desired communication rate to be $b=0.4$. In addition to the four algorithms compared for Problem 1, we include the parameter-based learning algorithm in~\eqref{eq:rate estimate}-\eqref{eq:bernouli estimate}. We show the results of the communication rate and the average cost in Figs.~\ref{fig:constrained_comm_rate} and~\ref{fig:constrained_aver_cost}. The empirical value of the communication rate and the average estimation error are computed in the same way as before. It can be seen that the four stochastic approximation-based algorithms have comparable performances in terms of the communication rate. Moreover, their empirical average estimation errors are comparable to that of the direct parameter learning method.

\begin{figure}[t]
	\centering
	\includegraphics[width=0.3\textwidth]{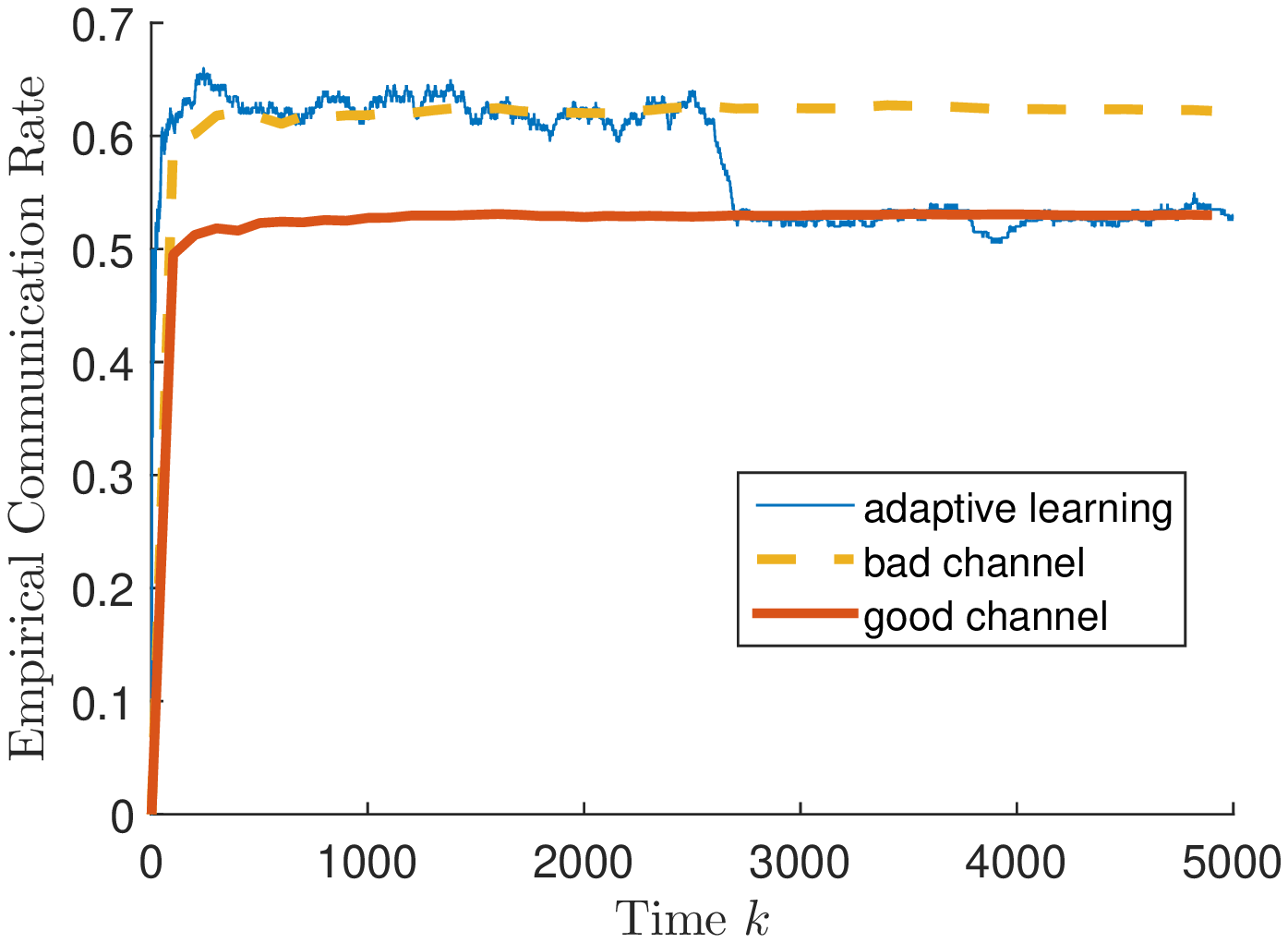}
	\caption{The learning method is adaptive to time-varying channel condition.}
	\label{fig:adaptive_learn_comm_rate}
	\includegraphics[width=0.3\textwidth]{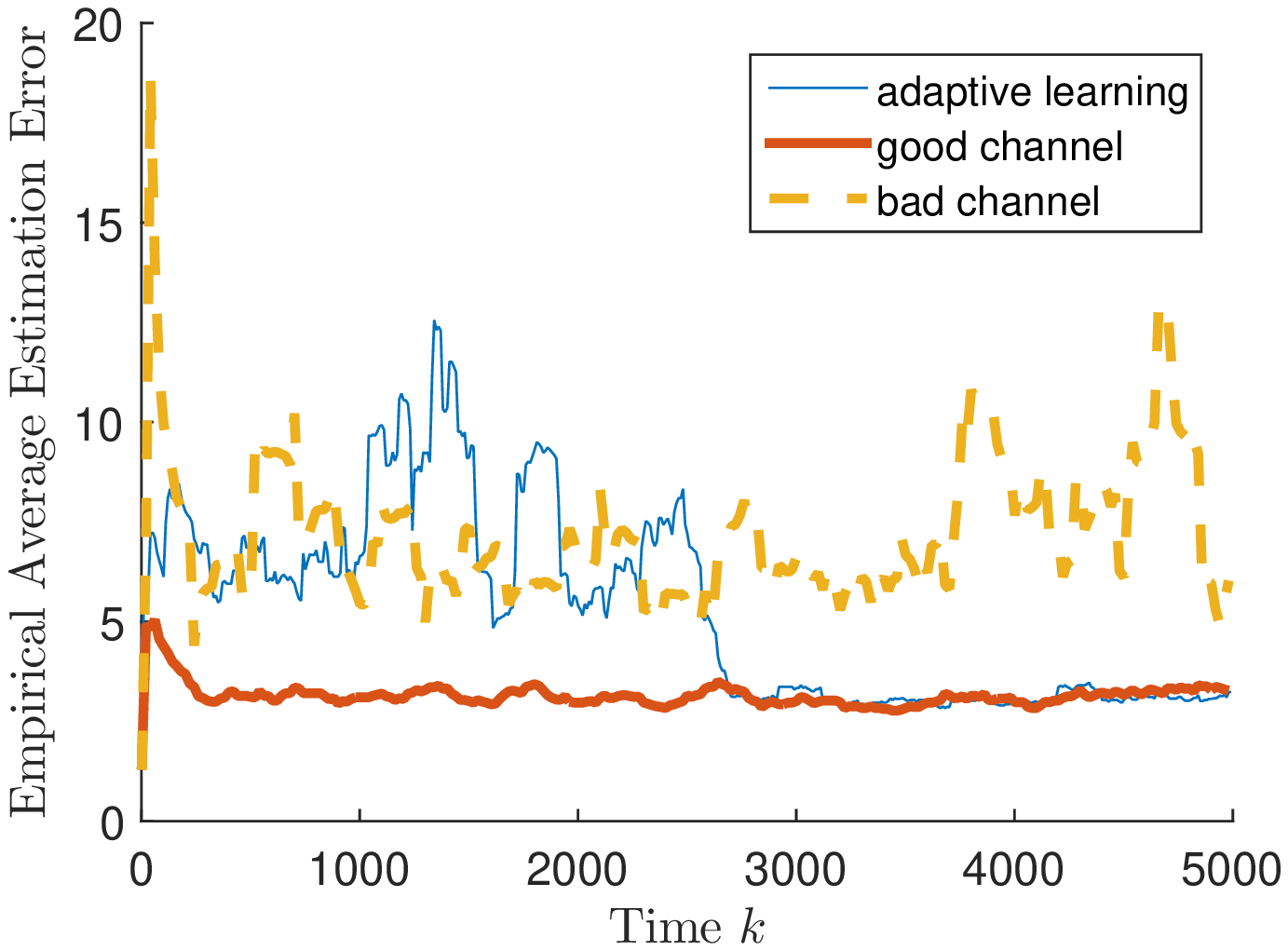}
	\caption{Average estimation error of three scheduling policies.}
	\label{fig:adaptive_learn_cost}
\end{figure}

We next illustrate the effectiveness of the learning method for a time-varying channel. We consider Problem 1 and set the communication cost to be $\lambda = 10$ per transmission. The channel condition is initially good with a successful transmission rate being $r_s=0.9$. At iteration time step $k=2500$, the successful transmission rate decreases to $r_s=0.6$. We compare the transient performance of the synchronous structured learning method with the performance under constantly ``good" or ``bad" conditions. Figs.~\ref{fig:adaptive_learn_comm_rate} and~\ref{fig:adaptive_learn_cost} show that the learning method is adaptive to time-varying channel conditions as the empirical communication rate and the empirical average estimation error converge to the optimal value. The computation of the empirical values of the communication rate and the average estimation error is computed using a sliding window to compute the values as
\begin{align*}
&\tilde{J}_{e,w}(k) =
\begin{cases}
\frac{1}{k+1} \sum_{t=0}^{k} c_e(\tau(t),a(t)),&~\text{if}~k<T_w,\\
\frac{1}{T_w} \sum_{t=k-T_w+1}^{k} c_e(\tau(t),a(t)),&~\text{if}~ k\geq T_w,
\end{cases}\\
&\tilde{J}_{r,w}(k) =
\begin{cases}
\frac{1}{k+1} \sum_{t=0}^{k} a(t),&~\text{if}~k<T_w,\\
\frac{1}{T_w} \sum_{t=k-T_w+1}^{k} a(t),&~\text{if}~k\geq T_w.
\end{cases}
\end{align*}
The solid blue lines are the adaptive learning method, while the solid red line and the dotted orange line are under ``good" and ``bad" channel conditions, respectively. Note that as the entropy of a Bernoulli random variable with mean being $0.6$ is greater than that with mean being $0.9$, the empirical average estimation error under bad channel has a greater fluctuation.

\begin{figure}[h]
	\centering
	\includegraphics[width=0.4\textwidth]{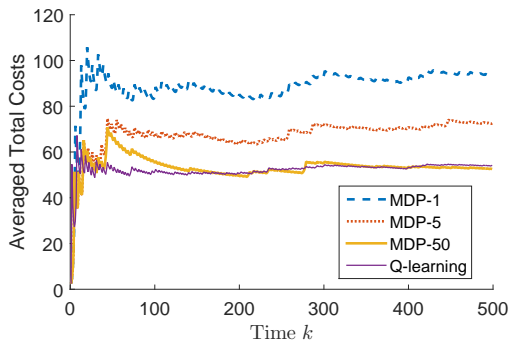}
	\caption{\noindent\textit{Performance comparison between the $Q$-learning and the parameter learning. The label ``MDP-x" stands for the parameter learning method with x allowable iterations at each time step.}}
	\label{fig: comparison of Q learning and adaptive control}
\end{figure}

We mentioned in the introduction that adaptive control methods can be computationally intense. We show how the $Q$-learning-based methods outperform direct parameter learning. In particular, we consider the remote estimation of the same dynamic process as previous examples for Problem 1. We simultaneously run the synchronous $Q$-learning with the parameter learning algorithm. In the parameter learning algorithm, we first estimate $r_s$ based on the history of transmission success and failures, and then calculate the optimal policy of the corresponding MDP using the relative value iteration. As the relative value iteration fails to converge within finite time, we forcefully stop the algorithm within $1$, $5$ and $50$ iterations. The time-averaged costs of each algorithm is presented in Fig.~\ref{fig: comparison of Q learning and adaptive control}. The label ``MDP-x" stands for the parameter learning method with x iterations at each time step.
If only one iteration is allowed for the MDP algorithm at each time step, the performance of the parameter learning is much worse than the $Q$-learning. If the number of iterations increases, the performance improves. The performance of the parameter learning is close to the $Q$-learning for $50$ iterations at each time step. The computation overhead of the $Q$-learning is equivalent to one iteration of the relative value iteration for MDP. In this particular example, it costs approximately $50$ times more computation resources for the parameter learning method to reach the same performance as that of the $Q$-learning.
\section{CONCLUSION}
We considered scheduling for remote state estimation under costly communication and constrained communication, respectively. By using dynamic programming, we established two frameworks to tackle the problems when the channel condition is known. We utilized these results to develop revised algorithms to improve the convergence of the standard asynchronous stochastic approximation algorithm. In addition, as the randomness of the state transition was observed to be independent of the state, we developed a simple synchronous algorithm for the costly communication problem. Although the stochastic approximation method can be used for the constrained communication problem, the parameter learning method possesses faster convergence speed and is easier to implement. For future work, the framework can be extended to a general channel such as a Markovian channel and scheduling multiple sensors.

\section*{Appendix}
\subsection{Proof of Lemma~\ref{lemma: sen condition for acoe}}
We adopt~\cite[Theorem 5.5.4]{hernandez1995discrete} to prove the result.
Denote the discounted total cost under a policy $f$ as
\begin{align*}
V_{\gamma}(\tau):= \sum_{k=0}^\infty \gamma^k\mathbb{E}[c(\tau(k),f(\tau(k)))|\tau(0)=\tau]
\end{align*}
and the infimum of $V_{\gamma}(\tau)$ as 
\begin{align*}
V_{\gamma}^\star(\tau) = \inf_{f\in\mathbb{F}} V_{\gamma}(\tau).
\end{align*}
In summary, the following conditions need to be verified.
\begin{enumerate}
	\item The one-stage cost $c(\tau,a)$ is continuous, nonnegative, and for any $r\in\mathbb{R}$ the set $\{a\in\mathbb{A}| c(\tau,a)<r\}$ is compact.
	\item The probability transition law $\mathtt{Pr}(\tau_+|\tau,a)$ is strongly continuous in $a$.
	\item\label{condition of boundedness} There exists a state $z\in\mathbb{S}$, a number $0<\underline{\gamma}<1$ and $\overline{M}\geq 0$ such that
	\begin{align*}
	(1-\gamma) V^\star_{\gamma}(z) \leq \overline{M},~\forall \tau\in\mathbb{S},~ \underline{\gamma}\leq\gamma<1.
	\end{align*}
	\item There exists a constant $\underline{M}\geq 0$ and a nonnegative function $u(\tau)$ on $\mathbb{S}$ such that
	\begin{align*}
	-\underline{M} \leq V^\star_{\gamma}(\tau)-V^\star_{\gamma}(z) \leq u(\tau),~\forall \tau\in\mathbb{S},~ \underline{\gamma}\leq\gamma<1,
	\end{align*}
	where the state $z$ is the same as in~(\ref{condition of boundedness}).
	\item The function $u(\tau)$ above is measurable and for any $\tau\in\mathbb{S}$ and $a\in\mathbb{A}$: $\sum_{\tau_+\mathbb{S}} u(\tau_+)\mathtt{Pr}(\tau_+ | \tau,a)<\infty$
	\item The sequence $\{ V^\star_{\gamma(n)}(\tau)-V^\star_{\gamma(n)}(z)\}$ is equicontinuous.
\end{enumerate}
The first two conditions hold as the actions set $\mathbb{A}$ is finite. According to Abelian Theorem~\cite[Lemma 5.3.1]{hernandez1995discrete}, we have
\begin{align*}
&\liminf_{T\to\infty} \frac{1}{T+1}\sum_{k=0}^{T} c_k \leq \liminf_{\gamma\uparrow 1} (1-\gamma) \sum_{k=0}^{\infty} \gamma^k c_k \\ \leq &\limsup_{\gamma\uparrow 1} (1-\gamma) \sum_{k=0}^{\infty} \gamma^k c_k \leq \limsup_{T\to\infty} \frac{1}{T+1}\sum_{k=0}^{T} c_k.
\end{align*}
As the average cost is bounded if we always set $a=1$, condition (3) holds for any state. We pick $z=0$. Let $T=\inf_{t\geq 0}\{t:\tau_t=z\}$. We can obtain
\begin{align*}
V_{\gamma}^\star(\tau) &\leq  \mathbb{E}[\sum_{t=0}^T \alpha^t c(\tau(t),1) |\tau_0=\tau] + \mathbb{E}[\alpha^T|\tau_0=\tau] V_{\gamma}^\star(z) \\
&= \mathbb{E}[\sum_{t=0}^T c(\tau(t),1) |\tau_0=\tau] + V_{\gamma}^\star(z)
\end{align*}
for any $\tau\neq z$. One can verify that
\begin{align*}
\mathbb{E}[\sum_{t=0}^T c(\tau(t),1) |\tau_0=\tau] = \sum_{t=0}^\infty r_s(1-r_s)^t c(\tau+t,1)
\end{align*}
is bounded for every $\tau$ as long as $(1-r_s)\rho(A)^2<1$. We can thus set $u(\tau) = \mathbb{E}[\sum_{t=0}^T c(\tau(t),1) |\tau_0=\tau]$ to satisfy condition (4). Since there are finite $\tau_+$ for a $\tau$ in $\mathtt{Pr}(\tau_+|\tau,a)$, condition (5) also holds. Condition (6) holds as the state space is discrete. Since the six conditions hold, we can use the vanish discount approach~\cite[Section 5.3]{hernandez1995discrete} to show the existence of an optimal stationary policy for the average cost problem.

\subsection{Proof of Lemma~\ref{lemma: monotonicity of V}}
Similar to the existence of an optimal stationary policy, the proof relies on a discounted cost setup for the same problem. For a constant $0<\gamma<1$, we want to minimize the discounted total cost $\sum_{k=0}^\infty \gamma^k\mathbb{E}[c(\tau(k),a(k))]$. The optimal policy satisfies the Bellman optimality equation for the discounted cost problem
\begin{align*}
V_{\gamma}(\tau)=\min_{a\in\mathbb{A}} \Big[ c(\tau,a)+\gamma\sum_{\tau+}V_{\gamma}(\tau_+)\mathrm{Pr}(\tau_+|\tau,a) \Big].
\end{align*}
Note that the right hand side of the discounted Bellman optimality equation is a mapping of $V_{\gamma}(\tau),\tau\in\mathbb{S}$. Define such mapping as the Bellman operator on $V_\gamma(\tau),\tau\in\mathbb{S}$ as
\begin{align*}
\mathcal{T} (V_{\gamma})= \min_{a\in\mathbb{A}} \Big[ c(\tau,a)+\gamma\sum_{\tau+}V_{\gamma}(\tau_+)\mathrm{Pr}(\tau_+|\tau,a) \Big].
\end{align*}
The discounted setup is considered here because the Bellman operator $\mathcal{T}_\gamma$ for the discounted cost problem is a contraction mapping w.r.t. to the later defined sup-norm. Since there is a unique fixed point for a contraction mapping iteration, which enables us to use an induction-based method to prove the monotonicity of the discounted value function. Moreover, as the six condition in Lemma~\ref{lemma: sen condition for acoe} hold, we have $V(\tau)=\lim_{\gamma\uparrow1} V_{\gamma}(\tau)$. In the following, we first show the contraction property of the Bellman operator. Then, we use induction-based method to show the monotonicity of the discounted value function.

\subsubsection{Contraction property of the Bellman operator}
We introduce the contraction mapping on a function space and conditions that guarantee the contraction property. Define $w:\mathbb{S}\to[1,\infty]$ as a weight function. For a real-value function $u$ on $\mathbb{S}$, define its $w$-norm as $\|u\|_w:=\sup_{\tau\in\mathbb{S}} |u(\tau)|/w(\tau)$. Let $\mathbb{B}_w(\mathbb{S})$ be the normed linear space of $w$-bounded functions on $\mathbb{S}$. If a mapping $\mathcal{T}$ from $\mathbb{B}_w(\mathbb{S})$ to $\mathbb{B}_w(X)$ is a contraction mapping w.r.t. the $w$-norm, we have
\begin{align*}
\|\mathcal{T}^k V_\gamma - V_\gamma^\star \|_w \leq \eta^k \|V_\gamma-V_\gamma^\star\|_w,~ V\in\mathbb{B}_w(\mathbb{S}),
\end{align*}
where $\eta<1$.

According to~\cite[Theorem 8.3.6]{hernandez1999further}, the Bellman operator $\mathcal{T}$ is a contraction mapping w.r.t. the $w$-norm if a set of conditions holds~\cite[Assumption 8.3.1, 8.3.2 and 8.3.3]{hernandez1999further}. Assumption 8.3.1 and 8.3.3 impose requirements on the compactness of the action space and continuity w.r.t. the actions, which holds straightforwardly in our setup. Assumption 8.3.2 requires that there exist constants $\overline{c}\geq0$ and $1 \leq \alpha \leq 1/\gamma$ and a weight function such that for every $\tau\in\mathbb{S}$: (1) $ \sup_{a\in\mathbb{A}} |c(\tau,a)|\leq \overline{c}w(\tau)$; (2) $ \sup_{a\in\mathbb{A}} \mathbb{E}[w(\tau_+)|\tau,a]\leq \alpha w(\tau)$. Based on these conditions, we have the following lemma, which is the same as the sufficient condition to guarantee the existence of a deterministic and stationary optimal scheduling policy.

\begin{lemma}
	If $\rho^2(A)(1-r_s)<1$, the Bellman operator $\mathcal{T}$ is a contraction mapping w.r.t. to the $w$-norm.
\end{lemma}

\begin{IEEEproof}
	It only suffices to verify the two conditions. Let $q>0$ satisfies $\overline{P} \preceq q\cdot I$ and $\Sigma_w \preceq q\cdot I$. According to~\eqref{eq: error covariance and holding time}, we can obtain that
	\begin{align*}
	P(\tau) \preceq q \sum_{t=0}^{\tau} A^t(A^{\top})^t,
	\end{align*}
	where $\preceq$ stands for the partial order in terms of the positive definite matrix. As both sides of the inequality are symmetric positive semidefinite matrices, we can obtain that
	\begin{align*}
	\Tr(P(\tau)) + \lambda \leq qn^2 \sum_{t=0}^{\tau} \sigma_1^{2}(A^t) +\lambda \leq q_{\sigma} \sigma_1^{2}(A^\tau)
	\end{align*}
	for some constant $q_{\sigma}>0$, where $\sigma_1(A)$ denotes the largest singular value of $A$. Note that $\sigma_1(A)$ is also the matrix $2$-norm, which satisfies
	\begin{align*}
	\lim_{\tau\to\infty} (\sigma_1(A^\tau))^{1/\tau} = \rho(A).
	\end{align*}
	Therefore,
	\begin{align*}
	\lim_{\tau\to\infty}\frac{(\Tr(P(\tau)) + \lambda)^{1/\tau}}{\rho^{2}(A)}
	=\lim_{\tau\to\infty}\frac{(\Tr(P(\tau)) + \lambda)^{1/\tau}}{(\sigma_1^{2}(A^\tau))^{1/\tau}} \leq q_{\sigma},
	\end{align*}
	which implies that there exists some $\overline{q}$ such that
	\begin{align*}
	\frac{\Tr(P(\tau)) + \lambda}{\rho^{2\tau}(A)}\leq \overline{q}
	\end{align*}
	for all $\tau\in\mathbb{S}$.
	
	Let
	\begin{align*}
	w(\tau) = 
	\begin{cases}
	\overline{q}\rho^{2L}(A),&\text{if~}\tau<L,\\
	\overline{q}\rho^{2\tau}(A),&\text{if~}\tau\geq L,
	\end{cases}
	\end{align*}
	We see that
	\begin{align*}
	\sup_{a\in\mathbb{A}} |c(\tau,a)| =  \Tr(P(\tau)) + \lambda \leq w(\tau),
	\end{align*}
	which verifies the first condition. For the second condition, note that we can restrict the policy in the following form: if $\tau\geq L$, $a=1$. When $\tau<L$, we have
	\begin{align*}
	\sup_{a\in\mathbb{A}} \mathbb{E}[w(\tau_+)|\tau,a] \leq \overline{q} \rho^{2L}(A) \leq w(\tau).
	\end{align*}
	When $\tau\geq L-1$, we have
	\begin{align}\label{eq:drift for w}
	\mathbb{E}[w(\tau_+)|\tau,1] &= (1-r_s) \rho^{2}(A) \overline{q}\rho^{2\tau}(A) + r_s \overline{q}\rho^{2L}(A)\nonumber\\
	&< \overline{q}\rho^{2}(A) r_s \overline{q}\rho^{2\tau}(A) + r_s \overline{q}\rho^{2L}(A) \nonumber\\
	&= w(\tau) + r_s \overline{q}\rho^{2L}(A).
	\end{align}
	According to~\cite[Remark 8.3.5]{hernandez1999further}, condition (2) can be replaced by the inequality~\eqref{eq:drift for w}. Thus condition (2) is also verified. This completes the proof.
\end{IEEEproof}

\subsubsection{Monotonicity of the discounted value function}
Since the Bellman operator $\mathcal{T}$ is proven to be a contraction mapping, we can prove the monotonicity of the $V_\gamma$ function through induction. Assume for all $\tau$ and $\tau'$, the lemma holds. Let $a=a^{\star}$ achieves the right hand side of discounted Bellman optimality equation for $\tau$, we can see that
\begin{align}\label{eq:achieve rhs of BE}
V_{\gamma}(\tau) &= c(\tau,a^\star) + \gamma\sum_{\tau_+}V_{\gamma}(\tau_+)\mathrm{Pr}(\tau_+|\tau,a^\star).
\end{align}
We can check that, for any $a\in\mathbb{A}$,
\begin{align*}
\sum_{t=0}^{\tau'_+}\mathrm{Pr}(t|\tau',a) \leq \sum_{t=0}^{\tau_+}\mathrm{Pr}(t|\tau,a),
\end{align*}
which leads to
\begin{align}\label{eq: stochastic monotone}
\sum_{\tau_+}V_{\gamma}(\tau_+)\mathrm{Pr}(\tau_+|\tau,a^\star) \geq \sum_{\tau'_+}V_{\gamma}(\tau'_+)\mathrm{Pr}(\tau'_+|\tau,a^\star).
\end{align}
Combining $c(\tau,a^\star)\geq c(\tau',a^\star)$,~\eqref{eq:achieve rhs of BE} and~\eqref{eq: stochastic monotone}, we obtain that
\begin{align*}
V_{\gamma}(\tau)=&c(\tau,a^\star) + \sum_{\tau_+}V_{\gamma}(\tau_+)\mathrm{Pr}(\tau_+|\tau,a^\star) - \mathcal{J}^\star\\
\geq & c(\tau',a^\star) + \sum_{\tau'_+}V_{\gamma}(\tau'_+)\mathrm{Pr}(\tau'_+|\tau,a^\star) - \mathcal{J}^\star \\
\geq & \min_{a\in\mathbb{A}} \Big[ c(\tau',a) + \sum_{\tau'_+}V_{\gamma}(\tau'_+)\mathrm{Pr}(\tau'_+|\tau,a) - \mathcal{J}^\star \Big] \\
=&V_{\gamma}(\tau').
\end{align*}

Since the (SEN) conditions holds as shown in the proof of Lemma~\ref{lemma: sen condition for acoe}, the monotonicity also holds for the average cost setup.

\subsection{Proof of Lemmas~\ref{lemma: monotonicity of Q} and~\ref{lemma: submodularity of Q}}
\subsubsection{Proof of Lemma~\ref{lemma: monotonicity of Q}}
	The monotonicity of the $Q$-factor holds because
	\begin{align*}
&Q(\tau,a)-Q(\tau',a)\\
\geq&\sum_{\tau_+}\min_{a\in\mathbb{A}} Q(\tau_+,a)\mathrm{Pr}(\tau_+|\tau,a) - \sum_{\tau'_+}\min_{a\in\mathbb{A}} Q(\tau'_+,a)\mathrm{Pr}(\tau'_+|\tau',a)\\
=&\sum_{\tau_+}V(\tau_+)\mathrm{Pr}(\tau_+|\tau,a)-\sum_{\tau'_+} V(\tau'_+)\mathrm{Pr}(\tau'_+|\tau',a)\geq 0.
\end{align*}
	This completes the proof.

\subsubsection{Proof of Lemma~\ref{lemma: submodularity of Q}}
	Since $a,a'\in\mathbb{A}=\{0,1\}$, let $a=1$ and $a'=0$.
	We can compute that
	\begin{align*}
	&Q(\tau,1)-Q(\tau,0)-Q(\tau',1)+Q(\tau',0)\\
	&=r_s\min_{a\in\mathbb{A}}Q(0,a)+(1-r_s)\min_{a\in\mathbb{A}}Q(\tau+1,a)-\min_{a\in\mathbb{A}}Q(\tau+1,a)\\
	&-r_s\min_{a\in\mathbb{A}}Q(0,a)-(1-r_s)\min_{a\in\mathbb{A}}Q(\tau'+1,a)+\min_{a\in\mathbb{A}}Q(\tau'+1,a)\\
	&=r_s[\min_{a\in\mathbb{A}}Q(\tau'+1,a)-\min_{a\in\mathbb{A}}Q(\tau+1,a)]\\
	&=r_s(V(\tau'+1)-V(\tau+1))\leq 0,
	\end{align*}
	which completes the proof.

\subsection{Proof of Theorem~\ref{theorem: threshold for costly comm}}
This argument is equivalent to that, if
	\begin{align*}
	Q(\tau,1)\leq Q(\tau,0),
	\end{align*}
	then
	\begin{align*}
	Q(\tau',1)\leq Q(\tau',0)
	\end{align*}
	for $\tau\leq\tau'$.
	
	Since $V(\tau+1)\leq V(\tau'+1)$, we obtain
	\begin{align*}
	Q(\tau',1) - Q(\tau',0) = &\lambda + r_sV(0)-r_sV(\tau'+1) \\
	\leq &\lambda + r_sV(0)-r_sV(\tau+1)\\
	=&Q(\tau,1)-Q(\tau,0)\leq0.
	\end{align*}
	This completes the proof.

\subsection{Proof of Lemma~\ref{lemma: necessary condition for optimal constrained policy}}
	As the problem is feasible in the sense that there exists $f\in\mathbb{F}$ such that $J_r(f)<b$. The optimal solution $(f^\star,\lambda^\star)$ to the saddle point problem should satisfy	
	\begin{align*}
	\lambda^\star(J_r(f^\star)-b)=0.
	\end{align*}
	If $\lambda^\star=0$, the optimal scheduling policy is always to transmit, which violates the constraint $J_r(f^\star)\leq b$. Therefore, $\lambda\neq 0$, and $J_r(f^\star)=b$ accordingly.

\subsection{Proof of Theorem~\ref{theorem: threshold for constrained comm}}
The proof relies on the following lemmas.
\begin{lemma}\label{lemma: costly value and comm price}
	The optimal threshold $\theta^\star$ for Problem 1 is monotonically increasing and piecewise constant w.r.t. $\lambda$.
\end{lemma}

\begin{lemma}\label{lemma: continuity of total cost}
	The optimal total cost $\inf_{\theta\in\mathbb{S}}J(\theta,\lambda) := \inf_{\theta\in\mathbb{S}} J_e(\theta) + \lambda J_r(\theta)$ is concave and continuous in $\lambda$.
\end{lemma}

\begin{lemma}\cite[Theorem 1, Sec 8.4]{luenberger1997optimization}\label{lemma:sufficiency}
	If there exists $\lambda^\star\geq0$ and a feasible policy $f^\star\in\mathbb{F}$ such that
	\begin{align*}
	J_e(f^\star)+\lambda^\star J_r(f^\star)\leq J_e(f)+\lambda^\star J_r(f),~\forall f\in\mathbb{F},
	\end{align*}
	then $f^\star$ is the solution to
	\begin{align*}
	\min_f \quad &J_e(f)\\
	s.t.\quad &J_r(f)\leq J_r(f^\star).
	\end{align*}
\end{lemma}

The first two lemmas relate the costly communication cost with the communication cost $\lambda$ while the last lemma proposes a sufficient a condition for the solution to a constrained optimization.

\begin{figure}[t]
	\centering
	\includegraphics[width=0.4\textwidth]{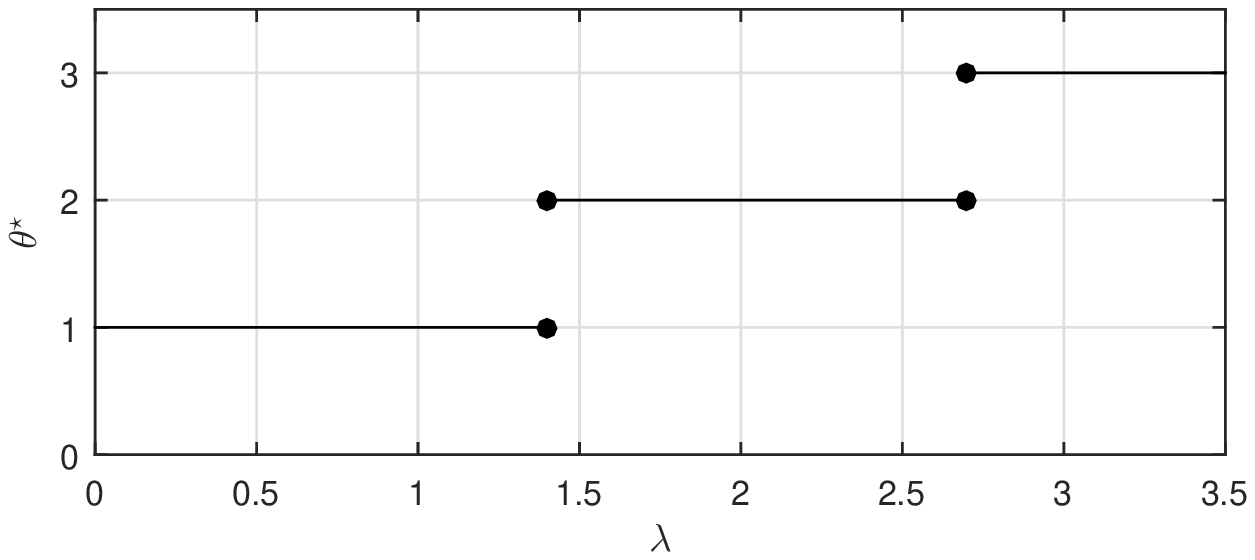}
	\caption{Optimal threshold $\theta^\star(\lambda)$ w.r.t. $\lambda$.}
	\label{fig:staircase}
\end{figure}

As $\theta$ is integer-valued, there must exist a sequence $\{\lambda_n\}_{n\in\mathbb{N}}$ such that the threshold policy with $\theta_n=n$ is optimal for $[\lambda_n,\lambda_{n+1})$ as shown in Fig.~\ref{fig:staircase}. In other words, the optimal total cost $\inf_{\theta\in\mathbb{S}} J(\theta,\lambda)$ is piecewise linear w.r.t. $\lambda$. The changing points of the slope of $\inf_{\theta\in\mathbb{S}} J(\theta,\lambda)$ w.r.t. $\lambda$ corresponds to the changing point of the optimal threshold $\theta^\star$ w.r.t. $\lambda$. At the discontinuity point, both $\theta$ and $\theta+1$ are optimal as they lead to the same total cost with fixed $\lambda$. This result is related to the continuity of $\inf_{\theta\in\mathbb{S}} J(\theta,\lambda)$ w.r.t. $\lambda$, which corresponds to Lemma~\ref{lemma: continuity of total cost}.

\subsubsection{Proof of Lemma~\ref{lemma: costly value and comm price}}
Note that
\begin{align*}
&[J(\theta,\lambda+1)-J(\theta,\lambda)] -  [J(\theta+1,\lambda+1)-J(\theta+1,\lambda)]\\
=& J_r(\theta)-J_r(\theta+1)\geq 0,
\end{align*}
for all $\theta$ and $\lambda$, which leads to 
\begin{align*}
J(\theta,\lambda+1)-J(\theta,\lambda) \geq J(\theta+1,\lambda+1)-J(\theta+1,\lambda)
\end{align*}
for all $\theta,\lambda$. Therefore, the total cost $J(\theta,\lambda)=J_e(\theta)+\lambda J_r(\theta)$ is submodular in $\theta$ and $\lambda$. From~\cite[Theorem 2.8.2]{topkis2011supermodularity}, $\theta^\star=\argmin_{\theta}J_e(\theta)+\lambda J_r(\theta)$ is increasing in $\lambda$.

\subsubsection{Proof of Lemma~\ref{lemma: continuity of total cost}}
As $J(\theta,\lambda)=J_e(\theta,\lambda)+\lambda J_r(\theta,\lambda)$ is affine (hence concave) in $\lambda$, the concavity of $\inf_{\theta\in\mathbb{S}}J(\theta,\lambda)$ follows directly.

Since $\inf_{\theta\in\mathbb{S}}J(\theta,\lambda)$ is piecewise linear, we only need to show that the continuity also holds at those changing points of the slope, i.e., $J(\theta_n,\lambda_{n+1})=J(\theta_n+1,\lambda_{n+1})$. We can observe that
\begin{enumerate}
	\item For any $\lambda_n \leq \lambda< \lambda_{n+1}$, we have $J(\theta_n,\lambda)\leq J(\theta_n+1,\lambda)$.
	\item For any $\lambda_{n+1} \leq \lambda< \lambda_{n+2}$, we have $J(\theta_n,\lambda)\geq J(\theta_n+1,\lambda)$.
\end{enumerate}
Since $J(\theta,\lambda)$ is continuous in $\lambda$, we obtain
\begin{align*}
\lim_{\lambda\to\lambda_{n+1}} J(\theta_n,\lambda)=\lim_{\lambda\to\lambda_{n+1}} J(\theta_n+1,\lambda).
\end{align*}
This completes the proof.

Now we are ready to prove the Theorem. According to Lemma~\ref{lemma:sufficiency}, it suffices to check that
\begin{enumerate}
	\item $J_r(f^\star)=b$. 
	\item There exists a $\lambda^\star$ such that $f^\star=\argmin_{f\in\mathbb{F}}J_e(f)+\lambda^\star J_r(f)$.
\end{enumerate}

The first condition follows straightforwardly from the description of $r_{\theta^\star}$ and $\theta^\star$ in the theorem. From Lemma~\ref{lemma: continuity of total cost}, we know that, for a given threshold $\theta^\star$, there exists a $\lambda^\star$ such that both $\theta^\star$ and $\theta^\star+1$ are optimal for the corresponding Problem 1. Therefore, any randomization between them is also optimal, which verifies the second condition. This completes the proof.
	
\subsection{Proof of Corollary~\ref{corollary: randomization policy}}
	For the given threshold $\theta$, we can calculate that 
	\begin{align*}
	\mathrm{Pr}(\tau\geq\theta)=\frac{1}{r_s\theta+1},
	\end{align*}
	which is $J_r(\theta)$. Since $\theta^\star$ should be the greatest integer that satisfies $J_r(\theta)\leq b$, we obtain the first formula.
	
	Given the threshold $\theta$ and the mean of the Bernoulli randomization $r_{\theta}$, the corresponding infinite-dimensional transition probability matrix of the Markov chain is given by
	\begin{align*}
	\begin{bmatrix}
	0 &1 \\
	\vdots & &\ddots\\
	0 & & &1 \\
	r_sr_{\theta} & & & &1-r_sr_{\theta}\\
	r_s & & & & &1-r_s\\
	\vdots & & & & & &\ddots
	\end{bmatrix},
	\end{align*}
	and the stationary distribution $\pi(\tau)$ for $\tau\in\mathbb{S}$ satisfies
	\begin{align*}
	\pi(i+1) &=\pi(i),\quad i=0,1,\dots,\theta-1,\\
	\pi(\theta+1) &= (1-r_sr_{\theta})\pi(\theta),\\
	\pi(\theta+1+i) &= (1-r_s) \pi(\theta+i), \quad  i=1,2,\dots
	\end{align*}
	As the stationary distribution is a probability distribution, we have
	\begin{align*}
	\sum_{i=0}^\infty \pi(i) = 1,
	\end{align*}
	which yields
	\begin{align*}
	\pi(i) = \frac{r_s}{r_s(\theta+1-r_{\theta})+1}, \quad i=0,1,\dots,\theta.
	\end{align*}
	In addition, the optimal policy should use the communication budget, which leads to the following balance equation
	\begin{align*}
	\pi(\theta^\star)\theta^\star + \pi(\theta^\star)(1-r_{\theta^\star}) = 1 -b,
	\end{align*}
	from which we obtain the second formula.

\subsection{Proof of Theorem~\ref{theorem: convergence of structured Q learning}}

According to~\cite{abounadi2001learning}, the standard asynchronous $Q$-learning converges, because the following two sets of conditions in~\cite{borkar2000ode} hold:
	\begin{enumerate}[]
		\item (A1) The function $h(\bm{Q}):=\nabla_{\bm{Q}} g(\bm{Q})$, the $(\tau,a)$-th element of which is $h_{(\tau,a)}(\bm{Q})=c(\tau,a)+\sum_{\tau'}\mathrm{Pr}(\tau'|\tau,a)\min_u Q(\tau',u) - Q(\tau,a) - Q(\tau_0,a_0)$, is Lipschitz continuous and the ODE $\dot{Q}=h(Q)$ is asymptotically stable at its equilibrium.
		\item (A2) The noise sequence $N_k$ is a martingale difference and is square integrable.
	\end{enumerate}
	
	Condition (A1) is related to the stability of an ODE, which is the average limit of the update equation~\eqref{eq:strutured learning primal} and~\eqref{eq:strutured learning dual}. Condition (A2) is related to the boundedness of the noise due to the noisy observation. In our structured $Q$-learning, the observation process is the same as the standard $Q$-learning. Therefore, it suffices to check the stability of the corresponding limit ODE.
	
	In the standard $Q$-learning, the asymptotic stability of the following ODE has been shown~\cite{abounadi2001learning}: $\dot{\bm{Q}} = h (\bm{Q})$. Consequently, we can obtain $\bm{Q}^\top h(\bm{Q}) =  \bm{Q}^\top \dot{\bm{Q}} <0$ for all points except the equilibrium $\hat{\bm{Q}}$ which satisfies $h(\hat{\bm{Q}})=0$. We need to check if the asymptotic stability also holds for the structured learning scheme.
	
	The ODE for the structured learning~\eqref{eq:strutured learning primal} and~\eqref{eq:strutured learning dual} is
	\begin{align}
	\dot{\bm{Q}} &= h (\bm{Q})+\bm{T}^{\top}\bm{\mu},\label{eq:ode with primal}\\
	\dot{\bm{\mu}}&= -\bm{T}\bm{Q}\label{eq:ode with dual}.
	\end{align}
	Consider the Lyapunov candidate as $V(\bm{Q},\bm{\mu}) = \frac{1}{2} \bm{Q}^{\top}\bm{Q} + \frac{1}{2} \bm{\mu}^{\top}\bm{\mu}$.
	We can derive that
	\begin{align*}
	\dot{V}= \bm{Q}^{\top}(h(\bm{Q})+\bm{T}^\top\bm{\mu}) -\bm{\mu}^\top\bm{T}\bm{Q} = \bm{Q}^\top h(\bm{Q}) <0,
	\end{align*}
	for all points except at $\hat{\bm{Q}}$. This proves that $\bm{Q}$ primal-dual scheme converges asymptotically to the solution of the Bellman optimality equation~\eqref{eq:bellman equation wrt q}.
	
\subsection{Proof of Theorem~\ref{theorem: convergence of the two-time scale}}

	The theorem can be proven by showing that the two-time scale iteration converges to the solution of the saddle point problem in~\eqref{eq:minmax constrained problem}. This is equivalent to $\lambda_{\infty}\in\argmax_{\lambda} J_e(f^\star)+\lambda J_r(f^\star)$ and $f^\star\in\argmin_f J_e(f^\star)+\lambda^\star J_r(f^\star)$, where $f^\star$ is the policy induced by $\bm{Q}_{\infty}(\cdot,\cdot)$.
	
	Similar to the stochastic approximation in one time scale, the two-time scale approach converges to the constrained communication problem if the two types of conditions in~\cite[Theorem 3.4]{bhatnagar2012stochastic} holds. The first type relates to the noise and the second type relates to the stability of the limit ODE. Based on analysis in Problem 1, the remaining task is to check the asymptotic stability of the ODE in the slower time scale. 
	
	A major difficulty lies in that the time average limit of the right hand side of ~\eqref{eq:price learning} is not an ODE but a differential inclusion as
	\begin{align*}
	\dot{\lambda} \in J_r(\lambda)-b
	\end{align*}
	as $J_r(\lambda)$ is discontinuous at countably many $\lambda$. Nevertheless, according to~\cite[Lemma 4.3]{borkar2005actor}, the limit ODE can be characterized by the following ODE instead
	\begin{align*}
	\dot{\lambda}(t) = \frac{\partial}{\partial \lambda} J^\star(\lambda(t)),
	\end{align*}
	where $J^\star(\lambda(t))=\inf_f J_e(f)+\lambda(t) J_r(f)$. The $\inf_f$ can be achieved as $\bm{Q}$ in the faster time scale converges according to previous analysis. The trajectory of $\lambda(t)$ is thus the solution to the following integral equation
	\begin{align*}
	\lambda(t) = \lambda(0) + \int_0^t J^\star(\lambda(s))  \, \mathrm{d} s.
	\end{align*}
	This interpretation conquers the discontinuity problem as the set of discontinuity has a zero measure. By the chain rule, the trajectory of the total cost $J^\star(t)$ satisfies
	\begin{align*}
	\dot{J^\star}(t) = \frac{\partial}{\partial \lambda} J^\star(\lambda)\cdot \dot{\lambda}(t) = |\frac{\partial}{\partial \lambda} J^\star(\lambda)|^2>0,
	\end{align*}
	for almost all $t$ except when $\lambda(t)\in\argmax J^\star(\cdot)$. This proves that $\lambda(t)$ converges to $\argmax J^\star(\cdot)$, i.e.,
	\begin{align*}
	J_e(f^\star)+\lambda_{\infty}(J_r(f^\star)-b)=J_e(f^\star)+\lambda^\star(J_r(f^\star)-b),
	\end{align*}
	where $\lambda^\star$ is the saddle point solution to~\eqref{eq:minmax constrained problem} and
	\begin{align*}
	f^\star(\tau) \in \argmin_{a} \Big\{ c(\tau,a) + \sum_{\tau'}\min_{u\in\mathbb{A}} \bm{Q}(\tau',u)\mathrm{Pr}(\tau'|\tau,a)-\mathcal{J}^\star \Big\}.
	\end{align*}
	This completes the proof.
	
\subsection{Proof of Theorem~\ref{theorem: convergence of the parameter learning}}
	For any sample path, $\lim_{k\to\infty} N_s(k)+N_f(k)=\infty$. By strong law of large numbers,  $\hat{r}_s(k)$ converges to $r_s$ almost surely. Therefore, thfe corresponding policy determined by $\hat{r}_s(k)$ asymptotically converges to the optimal policy. Note that the optimal estimation error is a continuous function of $r_s$. By the continuous mapping theorem~\cite[Theorem 3.2.4]{durrett2010probability}, the estimation error under the parameter learning policy converges almost surely.

\bibliographystyle{IEEETran}
\bibliography{strlearning}
\end{document}